\documentclass[aps,prd,twocolumn,showpacs,groupedaddress,superscriptaddress,nofootinbib,floatfix,preprintnumbers,longbibliography]{revtex4-1}
\usepackage{amssymb,amsmath,graphicx, color,mathrsfs}
\usepackage{physics}
\usepackage{bm}
\usepackage[tight]{subfigure}
\usepackage[export]{adjustbox}
\usepackage{braket}
\usepackage{svg}
\usepackage{mathtools}
\usepackage{bbm}
\usepackage{dsfont}
\usepackage{multirow}
\usepackage{rotating,booktabs}
\usepackage[verbose]{placeins}
\usepackage{color}
\usepackage{braket}
\usepackage{dsfont}

\usepackage{tikz,tkz-euclide}
\usetikzlibrary{arrows,calc,patterns}
\usetikzlibrary{decorations.markings}

\usepackage{quantikz}
\usepackage{url}
\usepackage{xcolor}
\usepackage{multirow}
\usepackage{rotating,booktabs}
\usepackage[verbose]{placeins}
\usepackage{ragged2e}
\usepackage[titletoc]{appendix}

\usepackage[colorlinks=true,citecolor=blue,linkcolor=red]{hyperref}
\usepackage{orcidlink}

\begin{document}

\dimen\footins=5\baselineskip\relax

\preprint{UMD-PP-026-06}

\title{Exponential-in-$N_c^2$ cost reduction of product-formula-based
\\
quantum simulations of quantum chromodynamics
}

\author{Zohreh Davoudi\,\orcidlink{0000-0002-7288-2810}}
\email{davoudi@umd.edu}
\affiliation{
Maryland Center for Fundamental Physics and Department of Physics, 
University of Maryland, College Park, MD 20742, USA}
\affiliation{Joint Center for Quantum Information and Computer Science,
NIST/University of Maryland, College Park, Maryland 20742}
\affiliation{The NSF Institute for Robust Quantum Simulation, University of Maryland, College Park, Maryland 20742, USA}

\author{Jesse R.~Stryker\,\orcidlink{0000-0002-4968-7988}}
\email{jstryker@lbl.gov}
\affiliation{
Physics Division, Lawrence Berkeley National Laboratory, Berkeley, CA 94720, USA}

\date{\today}

\begin{abstract}
Quantum algorithms for simulating quantum chromodynamics (QCD) have matured steadily since the pioneering work of Byrnes and Yamamoto [PRA 73, 022328 (2006)]. The most popular strategies for Hamiltonian simulation involve product-formula decompositions. However, the application of product-formula methods to SU($N_c$) lattice gauge theories by Byrnes and Yamamoto leads to $O(\Lambda^{8(N_c^2-1)})$ gate complexity per Trotter step, where $\Lambda$ is the bosonic cutoff in the electric (i.e., irreducible-representation) basis. A seminal work by Kan and Nam [arXiv:2107.12769 (2021)] significantly improves over such an undesirable cost and reports an $O\big(\Lambda \text{polylog}(\Lambda)\big)$ scaling, yet it still calls for an unrealistically large number of quantum gates. Here, we illuminate one of the reasons behind this high cost estimate and show that a factor of size $O(2^{4(N_c^2-1)})$ can be removed from the per-Trotter-step cost estimate by Kan and Nam. We specifically show that, by using methods developed in our past works [PRD 112, 014508 (2025); Quantum 7, 1213 (2023)], exponentiated-Hamiltonian decomposition---a necessary step in the application of product-formula algorithms---can be performed far more efficiently than previously thought. Our method reduces the T-gate cost estimate of QCD simulations using a second-order product formula by a factor of nearly $10^{14}$, independent of simulation parameters and sizes. Focusing on simulations in the electric basis, we further contrast our results with other methods: the local-multiplet basis approach of Ciavarella, Klco, and Savage [PRD 103, 094501 (2021)] and the near-optimal algorithm of Rhodes, Kreshchuk, and Pathak [PRX Quantum 5, 040347 (2024)]. This work highlights the importance of continued algorithmic improvement to bringing the quantum-simulation cost of QCD within reach of realistic quantum computers.
\end{abstract}

\maketitle

\tableofcontents

\section{\label{sec: Introduction}Introduction}

The field of quantum computation is fast evolving, and simulation of quantum many-body dynamics remains one of the key applications of quantum computers~\cite{Feynman:1981tf,Georgescu:2013oza,Daley:2022eja}. Quantum field theories, in particular, have long been identified as a target application. Quantum chromodynamics is especially relevant for many phenomenological predictions rooted in the Standard Model of particle physics~\cite{Gross:2022hyw,Achenbach:2023pba}; if successful, quantum computers will allow \emph{ab initio} simulations of real-time dynamics and far-from-equilibrium conditions in collider scattering events, the early universe, and matter in extreme environments~\cite{Klco:2021lap,Bauer:2022hpo,DiMeglio:2023nsa,Bauer:2023qgm,Beck:2023xhh,Funcke:2023jbq,Halimeh:2025vvp,Davoudi:2025kxb,Davoudi:2026ghz}. Essential to this program is the ability to simulate time evolution of QCD in Minkowski (as opposed to Euclidean) spacetime.

The most popular algorithms for simulating Schr\"odinger-picture time evolution on a universal quantum computer are based on product formulas (PFs)~\cite{Lloyd:1996aai}. Consider a Hamiltonian $H=\sum_{\gamma=1}^{\Gamma} H_\gamma $ with noncommuting $H_\gamma$. In quantum simulation, circuits for exactly implementing the time-evolution operator $e^{-i T H}$ (with $T\in \mathbb{R}$) are generally not available, so one must resort to an approximation method. PFs refer to approximating $e^{-i T H} $ by a sequence of \emph{subevolutions} of the form $e^{-i tH_\gamma}$ (with $|t|<|T|$). One must then develop circuits for the $e^{-i t H_\gamma}$ subevolutions. A commonly used PF is the second-order Trotter-Suzuki expansion~\cite{Trotter:1959ytf,Suzuki:1976be},
\begin{align}
    V_2(T;s) = &\left( \prod_{\gamma=1,2,...}^{\Gamma} e^{-i \frac{t}{2} H_\gamma} \prod_{\zeta=\Gamma, \Gamma-1, ...}^{1} e^{-i \frac{t}{2} H_{\zeta}} \right)^s,
    \label{eq:V2}
\end{align}
where $s\coloneq \tfrac{T}{t} \in \mathbb{Z}^+$. This form approximates the exact evolution with a spectral-norm error that is bounded as~\cite{Childs:2019hts}
\begin{align}
\left\|V_{2}(T;s)-e^{-iT H}\right\| &\leq 
\frac{\alpha_2\,T^{3}}{s^2}.
\label{eq:p2-formula}
\end{align}
Here, the commutator-error bound $\alpha_2$ is a function of the spectral norm of the second-order nested commutators of the Hamiltonian terms. 

PF algorithms take advantage of the locality that is usually present in physical theories, including in quantum field theories. Moreover, the PF algorithms do not intrinsically require ancillary qubits. Such features are especially appealing in noisy intermediate-scale quantum (NISQ) devices, but remain relevant for the fault-tolerant era of quantum computing too.

In this paper, we consider the application of PF algorithms to simulating real-time dynamics of gauge theories (including QCD) in the Kogut-Susskind~\cite{Kogut:1974ag} lattice formulation. Any PF algorithm of a quantum many-body Hamiltonian starts with a Trotterization prescription---a choice by which the collection of Hamiltonian summands $H_\gamma$ are identified. This \emph{splitting} is informed by whether or not $e^{-itH_\gamma}$ can be efficiently decomposed into a universal set of single- and two-qubit gates. If such a decomposition is not known, further splitting of terms may be required to obtain a set of $H_\gamma$ whose exponentiated forms can be implemented efficiently. This process often increases both the number of subevolutions to be implemented and the (bound on the) Trotter error. In particular, each individual subevolution in non-Abelian gauge theories may itself incur a high cost. It is therefore desirable to minimize the number of subevolutions, or equivalently $\Gamma$. Since lattice gauge theory (LGT) Hamiltonians are usually sums over local operators acting on a site, a link, or a plaquette, which are translationally invariant, we generally focus on one of such local operators at a time and use $\Gamma$ to refer to the subevolutions associated with that particular operator.

The most straightforward, but often inefficient, approach to Trotterized time evolution is \emph{Pauli decomposition}. 
Pauli decomposition entails determining the expression of a Hamiltonian in the basis of Pauli operators: $H=\sum_\gamma c_\gamma H_\gamma$, where each $H_\gamma$ is a Pauli operator and only nonzero $c_\gamma$ are included in the sum.
Here, we use \emph{Pauli operator} to refer to an arbitrary tensor product of the single-qubit Pauli operators
\begin{align}
    X=\ket{0}\!\!\bra{1}+\ket{1}\!\!\bra{0}, \ Z=\ket{0}\!\!\bra{0}-\ket{1}\!\!\bra{1}, \
    Y=i X Z,
\end{align}
and $\mathds{1}=\ket{0}\!\!\bra{0}+\ket{1}\!\!\bra{1}$.
Unitary evolution for an $H_\gamma$ that is a Pauli operator can be readily decomposed into elementary gates~\cite{Nielsen:2012yss}; however, for a Hamiltonian involving $\eta$ qubits, there may be as many as $4^\eta$ Pauli operators required for the decomposition.

Pauli decomposition is the underlying method in a pioneering paper by Byrnes and Yamamoto~\cite{Byrnes:2005qx} on quantum computing LGTs. They first write the Hamiltonian of a Yang-Mills LGT in terms of various number and ladder operators that are defined in the (discrete) irreducible-representation---or \emph{irrep}, also referred to as \emph{electric}---basis with truncation $\Lambda$. If the local irrep quantum numbers are encoded into qubits using a unary map, as proposed in Ref.~\cite{Byrnes:2005qx}, then the corresponding Pauli decomposition is simple but the qubit-register size is unnecessarily large. Using a binary map, the local irrep quantum numbers can be encoded exponentially more efficiently, but the Pauli decomposition is more involved. The number of Pauli terms in any case will be unrealistically large, and at worse scales as $O(\Lambda^{8(N_c^2-1)})$ for SU($N_c$) theories, which is inefficient. Furthermore, the Pauli terms often do not commute, which causes significant commutator error and drives up the number of Trotter steps, $s$, required to guarantee a fixed, low error [per Eq.~\eqref{eq:p2-formula}].

To remedy this situation, a seminal paper by Kan and Nam~\cite{Kan:2021xfc} adopts an earlier technique of Shaw et al.~\cite{Shaw:2020udc} to avoid a naive Pauli decomposition of the Hamiltonian terms, achieving a gate complexity that scales with $\Lambda$ as $O\big(\Lambda \, \text{polylog}(\Lambda)\big)$. 
The key idea from Ref.~\cite{Shaw:2020udc} is the artificial splitting of a bosonic ladder operator into two terms that we name as \emph{even} and \emph{odd}.
Kan and Nam apply the even-odd splitting from Ref.~\cite{Shaw:2020udc} to all ladder operators in the LGT Hamiltonian, multiplying them all out to obtain the summands for Trotterization.
This method for splitting the Hamiltonian leads to subevolutions on different subspaces that can be circuitized in closed form. In the U(1) theory in (1+1)D, the even-odd splitting leads to only a doubling of the interaction terms of the Hamiltonian, i.e., the fermion hopping terms. But when the method is carried over from U(1) to SU(2) to SU(3), and from (1+1)D to (2+1)D to (3+1)D, the abundance of ladder operators that all undergo splitting leads to an apparent need for $\sim 10^{15}$ subevolutions per magnetic (plaquette) term in the Hamiltonian~\cite{Kan:2021xfc}. A Trotter decomposition of this sort is still considered ``efficient'', from the point of view of computational complexity classes, because the large but fixed number of local subevolutions is independent of tunable simulation parameters (such as volume or time).
The staggering multiplicative cost that comes with $\sim 10^{15}$ subevolutions, however, presents a roadblock to implementations on near-term and potentially also far-term hardware, rendering the decomposition of Ref.~\cite{Kan:2021xfc} inefficient in the colloquial sense of the word.

We have found in Refs.~\cite{Stryker:2021asy,Davoudi:2022xmb} a different strategy to subdivide operators in the Hamiltonian. The strategies in Ref.~\cite{Kan:2021xfc} and Refs.~\cite{Stryker:2021asy,Davoudi:2022xmb} can all be understood as finding splittings for local Hamiltonian operators, $H\to \sum_{\gamma} H_\gamma$, wherein each $H_\gamma$ is block-diagonalized into blocks no larger than $2\times2$. Summands of that form are sought because if $H_\gamma$ only generates rotations in two-dimensional subspaces, analytically diagonalizing it is potentially tractable. However, one summand $H_\gamma$ could generate rotations in precisely one two-dimensional subspace, or in many disjoint two-dimensional subspaces simultaneously. If one can find diagonalizable $H_\gamma$ that generate rotations in many two-dimensional subspaces simultaneously, then fewer overall summands are needed.

References \cite{Stryker:2021asy,Davoudi:2022xmb} are rooted in the observation that if $H_\gamma$ involves a definite tensor product of ladder operators, then an even-odd splitting of just \emph{one} ladder-operator factor should suffice to yield new summands $H_{\gamma,\gamma'}$ (with $H_\gamma=\sum_{\gamma'}H_{\gamma,\gamma'}$) that generate rotations in two-dimensional subspaces and whose diagonalizations are tractable.
Reference~\cite{Stryker:2021asy} proposes a partial-diagonalization procedure based on systematically applying unitary transformations---which can be geometrically interpreted as shearing---until the transformed version of $H_\gamma$, say $\tilde{H}_\gamma$, involves exactly one ladder-operator factor. One then does a single even-odd splitting on the residual ladder operator, $\tilde{H}_\gamma\to\tilde{H}^\mathrm{(e)}_\gamma+\tilde{H}^\mathrm{(o)}_\gamma$, and separately diagonalizes $\tilde{H}^\mathrm{(e)}_\gamma$ and $\tilde{H}^\mathrm{(o)}_\gamma$ for circuitization. Reference~\cite{Davoudi:2022xmb} instead introduces at the start an even-odd splitting to any one of the bosonic ladder operators: $H_\gamma\to H^\mathrm{(e)}_\gamma + H^\mathrm{(o)}_\gamma$. With the help of an ancilla qubit, one applies certain unitary transformations---based on singular value decompositions (SVDs)---that have the effect of diagonalizing either $H^\mathrm{(e)}_\gamma$ or $H^\mathrm{(o)}_\gamma$ for circuitization. Later on, we discuss the Hamiltonian splitting assuming the SVD-based procedure \cite{Davoudi:2022xmb}, although one should be equally able to apply the shearing procedure \cite{Stryker:2021asy}.
We also provide a graph-theoretic interpretation to lend some intuition for our splitting scheme.

As will be shown in the paper, the strategy based on a single even-odd splitting reduces the number of subevolutions to be implemented per Trotter step in the SU($N_c$) LGT by a factor of $O(2^{4(N_c^2-1)})$ compared to Ref.~\cite{Kan:2021xfc}. For QCD, the reduction is a factor of $\sim10^9$ in the number of subevolutions per Trotter step, and a factor of nearly $10^{14}$ in the total simulation cost estimate of a second-order PF implementation. The reduction holds regardless of the cutoff value, system size, or other parameter values.

Our work demonstrates that continuous algorithmic improvements are vital to realistic resource estimation of gauge-theory simulations toward the fault-tolerant era of quantum computing, and they may decide feasibility of QCD simulations. To put our work in the broader context, we also contrast our algorithm with two other leading algorithms for simulating time dynamics of the Kogut-Susskind SU($N_c$) LGTs in the irrep basis: the post-Trotter, near-optimal algorithm of Rhodes, Kreshchuk, and Pathak~\cite{Rhodes:2024zbr}, and the local-multiplet basis approach of Ciavarella, Klco, and Savage~\cite{Ciavarella:2021nmj}. We further motivate a full end-to-end algorithmic analysis within different frameworks and algorithms presented to date~\cite{Lamm:2019bik,Haase:2020kaj,Murairi:2022zdg,Lamm:2024jnl,Balaji:2025afl,Halimeh:2025ivn,Ciavarella:2024fzw,Froland:2025bqf,Zache:2023dko,Rhodes:2026atz,Webb-Mack:2026bkg}.

This paper is organized as follows. In Secs.~\ref{sec:U(1)}, \ref{sec:SU(2)}, and \ref{sec:SU(3)}, we explain how our strategy is applied to the U(1), SU(2), and SU(3) LGTs, respectively, in any number of dimensions. Here, only the fermion-gauge interaction (hopping) and magnetic (plaquette) Hamiltonians are of importance given their off-diagonal nature in the irrep basis. Table \ref{tab: summand counts} compiles a summary of the number of subevolutions to be implemented within our approach compared with Ref.~\cite{Kan:2021xfc}. The overall cost saving for QCD simulations using a second-order PF is further provided. Section~\ref{sec:others} includes a discussion of other leading algorithmic approaches to the problem of quantum simulating QCD dynamics. In Sec.~\ref{sec:outlook}, we present our conclusions and offer an outlook.
\begin{table*}[t!]
    \centering
    \begin{tabular}{l r r r}
          Hamiltonian term & 
          Ref.~\cite{Kan:2021xfc} (quoted) & Ref.~\cite{Kan:2021xfc} (tightened)  & $\ $This work \\
        \hline
        U(1) hopping & 2 & 2 & 1 \\
        U(1) plaquette & 16 & 16 & 2 \\
        SU(2) hopping & 64 & 64 & 8 \\
        SU(2) plaquette & 1,048,576 & 524,288 & 256 \\
        SU(3) hopping & 13,872 & 9,248 & 75 \\
        SU(3) plaquette & 1,470,021,852,266,496 & $\ \ $ 290,374,686,867,456 & 1,062,882
    \end{tabular}
    \caption{The count of summands, $\Gamma$, in the product-formula decomposition of different operators in the Kogut-Susskind Hamiltonian, obtained using the two splitting methods. The irrep basis is assumed, and the chosen operators are those that are off-diagonal in this basis. The second column states the precise upper bound reported in Ref.~\cite{Kan:2021xfc}, while the third column reflects a careful recounting of strictly necessary summands within the same splitting algorithm. The final column is the number of summands obtained by the proposed splitting method in this work.}
    \label{tab: summand counts}
\end{table*}
%

\section{U(1) lattice gauge theory
\label{sec:U(1)}
}
It is instructive to first explain how our method improves the PF-based algorithm cost for more elementary lattice gauge theories, before generalizing to QCD. Consider the U(1) LGT coupled to staggered fermions in $d+1$ dimensions. An individual U(1) gauge link is characterized by one quantum number and has an on-link Hilbert space spanned by states of the form $\ket{\epsilon}$, where each electric-field value $\epsilon \in \mathbb{Z}$ corresponds to a distinct irrep of U(1). This basis, i.e., the irrep basis, is to be contrasted with its conjugate group-element basis $\ket{ \theta }$, wherein states are parametrized by a continuous coordinate $\theta$ on the U(1) manifold.

In the irrep basis, the link operator is defined by
\begin{align}
    U \ket{\epsilon} &= \ket{\epsilon-1},
    \label{eq:U(1)-link}
\end{align}
or equivalently $U  = \sum_\epsilon \ket{\epsilon - 1}\!\!\bra{\epsilon}$. Thus, $U$ is formally an infinite-dimensional shift matrix; it can be thought of as a matrix with ones on the superdiagonal and zeroes everywhere else. In practice, one often truncates the Hilbert space to some maximum value of the electric-field magnitude: $|\epsilon| \leq \Lambda$, rendering $U$ finite-dimensional.

To Trotterize terms in the Hamiltonian involving link operators, we turn to the even-odd splitting procedure as it was done in Ref.~\cite{Davoudi:2022xmb}. This algorithm is motivated by the fact that the link operator is a sparse matrix that has only superdiagonal elements, hence one can easily identify unitary matrices $\mathscr{V}$ and $\mathscr{W}$ such that $\mathscr{V}^\dagger U \mathscr{W}$ is properly diagonal. We supplement our explanation with a graph-theoretic interpretation, following a description presented in Ref.~\cite{Stryker:2021asy} for the U(1) case.

\subsection{Hopping terms}
To demonstrate the algorithm that implements each exponentiated Hamiltonian term in a Trotter step, we begin with the hopping terms---also known as the fermion--gauge-boson interaction---in the U(1) LGT. Each hopping term takes the form
\begin{equation}
    H_{\mathrm{hop}} = \psi_{\mathtt{x}}^\dagger U_{\mathtt{p}} \psi_{\mathtt{y}} + {\rm H.c.},
    \label{eq:U(1)hopping-psi}
\end{equation}
up to a constant multiplicative factor. Here, $\psi^\dagger$ and $\psi$ are the single-component, staggered-fermion creation and annihilation operators, respectively, $\mathtt{x}$ and $\mathtt{y}$ are any two nearest-neighbor sites on a Cartesian lattice, and $\mathtt{p}$ denotes the link connecting them. We often use the site or link labels to also refer to the registers of qubits associated with their respective degrees of freedom. Thus, $\mathtt{x}$ and $\mathtt{y}$ are each one-qubit registers, while $\mathtt{p}$ is an $\eta \coloneq \lceil \log_2(2\Lambda+1)\rceil$ qubit register. Applying a Jordan-Wigner transformation, the hopping term can be written in terms of qubit operators acting on the $\mathtt{x}$ and $\mathtt{y}$ registers. For the simplified case of a (1+1)D lattice, one has
\begin{equation}
    H_{\mathrm{hop}} = \ket{0}\!\!\bra{1}_\mathtt{x} \ket{1}\!\!\bra{0}_\mathtt{y} U_\mathtt{p} + {\rm H.c.}
    \label{eq:U(1)hopping}
\end{equation}
In higher dimensions, the Jordan-Wigner transformation can leave behind additional Pauli-$Z$ operator-factors on the hopping terms (so-called Pauli-$Z$ strings); these Pauli-$Z$ strings are already diagonal in the computational basis, and as such their presence or absence does not make any difference in the splitting algorithm that follows. The goal is to implement $e^{-itH_\text{hop}}$ using known circuit elements. 

First, we define $\mathscr{A} \coloneq \ket{1}\!\!\bra{0}_\mathtt{y} U_\mathtt{p}$ and note that $\mathscr{A}^2 = 0$. 
Second, we observe that $\mathscr{A}$ is accompanied in $H_\text{hop}$ by the operator-factor $\ket{0}\!\!\bra{1}_\mathtt{x}$.\footnote{From the point of view of the algorithm, the structural similarity of $\ket{0}\!\!\bra{1}_{\mathtt{x}}$ to the factor $\ket{1}\!\!\bra{0}_{\mathtt{y}}$ in $\mathscr{A}$ is merely a coincidence. The key property of $\mathscr{A}$ is simply $\mathscr{A}^2=0$. Later examples will have different choices of $\mathscr{A}$ that square to zero by other means.} The squaring-to-zero property of $\mathscr{A}$ allows for a diagonalization of $\ket{0}\!\!\bra{1}_\mathtt{x}\mathscr{A}$ plus its Hermitian conjugate, as shown in Ref.~\cite{Davoudi:2022xmb}. To proceed, one finds an SVD of $\mathscr{A}$, i.e., choices of unitary $\mathscr{V}$ and $\mathscr{W}$ such that $\mathscr{A}=\mathscr{V} \mathcal{S} \mathscr{W}^\dagger$, with $\mathcal{S}$ being a diagonal and positive semi-definite square matrix. A solution is simple to obtain:
\begin{align}
&\mathscr{V}= X_\mathtt{y} \lambda^-_\mathtt{p},\\
&\mathscr{W} = \mathds{1}_\mathtt{y} \mathds{1}_\mathtt{p},\\
&\mathcal{S}=\ket{0}\!\!\bra{0}_\mathtt{y} \mathcal{D}_\mathtt{p}.
\end{align}
Here, $\lambda_\mathtt{p}^{\pm}=\sum_\epsilon \ket{\epsilon}\!\!\bra{\epsilon\mp 1}_\mathtt{p}$ and $ \mathcal{D}_\mathtt{p} \coloneq \sum_\epsilon \ket{\epsilon}\!\!\bra{\epsilon}_\mathtt{p}$. With $\mathtt{p}$ truncated to a finite number of $\eta$ qubits, one should ``wrap'' the $\lambda_{\mathtt{p}}^{\pm}$ shift operators by defining $\ket{\epsilon_{\mathrm{max}}+1}\coloneq\ket{\epsilon_{\mathrm{min}}}$ and $\ket{\epsilon_{\mathrm{min}}-1}\coloneq\ket{\epsilon_{\mathrm{max}}}$, ensuring the unitarity of $\lambda_{\mathtt{p}}^{\pm}$.\footnote{The ``wrapping'' of $\lambda_{\mathtt{p}}^{\pm}$ operators at the lower and upper cutoffs potentially allows for an unphysical mixing between $\ket{\epsilon_{\mathrm{max}}}$ and $\ket{\epsilon_{\mathrm{min}}}$; to prevent this mixing, one can modify the definition of $\mathcal{D}_{\mathtt{p}}$~\cite{Stryker:2021asy}.}

With the above SVD, the next step is to introduce the unitary
\begin{align}
  \mathscr{U}&=\mathsf{H}_\mathtt{x} \big(\ket{0}\!\!\bra{0}_\mathtt{x}  \mathscr{V}^\dagger+\ket{1}\!\!\bra{1}_\mathtt{x}  \mathscr{W}^\dagger\big),
  \label{eq:unitaryForSVD}
\end{align}
where $\mathsf{H}_{\mathtt{x}}=(X_{\mathtt{x}}+Z_{\mathtt{x}})/\sqrt{2}$ denotes the single-qubit Hadamard gate for site $\mathtt{x}$. It it is straightforward to verify that the Hamiltonian in Eq.~\eqref{eq:U(1)hopping} has the diagonalization
\begin{equation}
\mathscr{U} H_{\rm hop} \mathscr{U}^\dagger = Z_\mathtt{x}\mathcal{S} \, .
\end{equation}
On a quantum computer, the $\mathscr{U}$ operation involves controlled operations, basic addition primitives, and Pauli gates. Details about these known operations can be found in, e.g., Ref.~\cite{Davoudi:2022xmb}. Overall, the circuit for implementing $e^{-itH_{\rm hop}}=\mathscr{U}^\dagger e^{-it Z_\mathtt{x} \ket{0}\!\!\bra{0}_\mathtt{y} \mathcal{D}_\mathtt{p}} \mathscr{U}$ can be implemented efficiently, i.e., with a number of gates that scales as a low-degree polynomial in $\eta$. Incidentally, the decomposition for the U(1) hopping term presented in this example is exact and does not require any splitting to be implemented. Our count of $\Gamma=1$ is reported in the last column of Table \ref{tab: summand counts}.

In contrast, the algorithm of Ref.~\cite{Shaw:2020udc} (adopted by Ref.~\cite{Kan:2021xfc}) splits a single link operator into a sum of two operators on $\mathtt{p}$: one that is nonzero when applied to even-numbered basis states and one that is nonzero when applied to odd-numbered basis states.
Concretely, 
\begin{align}
U_\mathtt{p}= U_\mathtt{p} \mathcal{E}_\mathtt{p}+U_\mathtt{p} \mathcal{O}_\mathtt{p},
\label{eq:split}
\end{align}
where $\mathcal{E}_\mathtt{p}= \sum_k \ket{2k}\!\!\bra{2k}_\mathtt{p}$, $\mathcal{O}_\mathtt{p}= \sum_k \ket{2k+1}\!\!\bra{2k+1}_\mathtt{p}$, and $\mathcal{E}_\mathtt{p}+\mathcal{O}_\mathtt{p} = \mathds{1} $. With this splitting, $\psi_{\mathtt{x}}^\dagger U_\mathtt{p} \mathcal{E}_\mathtt{p} \psi_{\mathtt{y}} + {\rm H.c.}$ and $\psi_{\mathtt{x}}^\dagger U_\mathtt{p} \mathcal{O}_\mathtt{p} \psi_{\mathtt{y}} + {\rm H.c.}$ can each be diagonalized following a procedure similar to what is presented above. This split---which turns out to be unnecessary for diagonalizing the U(1) hopping term, as shown above---leads to twice the number of simulatable terms and introduces a new source of Trotter error. While the cost increase is insignificant in this example, dramatic cost amplification can arise from such unnecessary splittings in the more complex cases, as will become clear shortly.

\subsection{Plaquette terms}
Next, we demonstrate how our algorithm works for the plaquette Hamiltonian in the U(1) case. Consider the single-plaquette Hamiltonian
\begin{align}
    H_{\mathrm{plaq}} &= U_\mathtt{p} U_\mathtt{q} U_\mathtt{s}^\dagger U_\mathtt{t}^\dagger + \mathrm{H.c.} 
    \label{eq:U(1)-plaq}
\end{align}
(up to a constant multiplicative factor), where $\mathtt{p}$, $\mathtt{q}$, $\mathtt{s}$, and $\mathtt{t}$ label four independent but structurally identical gauge links. Due to the absence of fermionic creation and annihilation operators, which previously provided us with a nilpotent operator, the plaquette operator does not square to zero and is not a candidate for the $\mathscr{A}$ operator. To proceed, we split one of the link operators into a sum of two operators according to Eq.~\eqref{eq:split}. With this splitting, the plaquette Hamiltonian can be written as
\begin{eqnarray}
     H_{\mathrm{plaq}}=H_{\mathrm{plaq}}^{(\text{e})}+H_{\mathrm{plaq}}^{(\text{o})},
     \label{eq:even-odd-H}
\end{eqnarray}
with
\begin{subequations}
\begin{align}
     &H_{\mathrm{plaq}}^{(\text{e})} = \mathscr{A}^{(\text{e})} + {\rm H.c.}, ~ \mathscr{A}^{(\text{e})}  \coloneq
     ( U_\mathtt{p} \mathcal{E}_\mathtt{p}) U_\mathtt{q} U_\mathtt{s}^\dagger U_\mathtt{t}^\dagger ,
     \label{eq:Hplaqe}
     \\
     &H_{\mathrm{plaq}}^{(\text{o})} = \mathscr{A}^{(\text{o})} + {\rm H.c.}, ~ \mathscr{A}^{(\text{o})} \coloneq
     (U_\mathtt{p}\mathcal{O}_\mathtt{p}) U_\mathtt{q} U_\mathtt{s}^\dagger U_\mathtt{t}^\dagger.
     \label{eq:Hplaqo}
\end{align}
\end{subequations}

Focusing on $H_{\mathrm{plaq}}^{(\text{e})}$, one first notes that $(\mathscr{A}^{(\text{e})})^2 = 0$, making it a candidate for the SVD algorithm of Ref.~\cite{Davoudi:2022xmb}. One way to diagonalize $\mathscr{A}^{(\text{e})}$ plus its Hermitian conjugate using an SVD is to enlarge the Hilbert space with one ancillary qubit, which is prepared in the computational-zero state, $\ket{0}_{\rm anc}$~\cite{Davoudi:2022xmb}. Here the ancilla ensures that a reversible unitary operation on $\mathscr{A}^{(\text{e})}$, whose net effect on the main qubits is an SVD of $\mathscr{A}^{(\text{e})}$, can be straightforwardly found, as shown below. Explicitly, one first defines a unitary operator $\mathscr{P}$ which projects $\ket{0}_\text{anc}$ to $\ket{0}_\text{anc}$ if the state of the $\mathtt{p}$ link is in the nullspace of $\mathscr{A}^{(\text{e})}$, and to $\ket{1}_\text{anc}$ otherwise. A valid choice for $\mathscr{P}^{(\text{e})}$ is
\begin{align}
    \mathscr{P}^{(\text{e})} &= X_{\rm anc}\mathcal{E}_\mathtt{p} + \mathcal{O}_\mathtt{p} .
\end{align}
The effect of $\mathscr{P}^{(\text{e})}$ is
\begin{align}
    \mathscr{P}^{(\text{e})} ( H_{\mathrm{plaq}}^{(\text{e})} \otimes & \ket{0}\!\!\bra{0}_{\rm anc} ) \mathscr{P}^{{(\text{e})}\,\dagger} 
    \nonumber\\
    &= ( U_\mathtt{p} \mathcal{E}_\mathtt{p}) U_\mathtt{q} U_\mathtt{s}^\dagger U_\mathtt{t}^\dagger \ket{0}\!\!\bra{1}_{\rm anc}
    + \mathrm{H.c.},
\end{align}
which yields a form similar to Eq.~\eqref{eq:U(1)hopping} and is suitable for the application of the SVD algorithm. We choose the decomposition $\mathscr{A}^{(\text{e})} = \mathscr{V}^{(\text{e})} \mathcal{S}^{(\text{e})} \mathscr{W}^{{(e)}\,\dagger}$ with
\begin{align}
    & \mathscr{V}^{(\text{e})} = \lambda^{-}_\mathtt{p} \lambda^{-}_\mathtt{q} \mathds{1}_\mathtt{s} \mathds{1}_\mathtt{t},\\
    & \mathscr{W}^{(\text{e})} = \mathds{1}_\mathtt{p} \mathds{1}_\mathtt{q} \lambda^{-}_\mathtt{s} \lambda^{-}_\mathtt{t}, \\
    & \mathcal{S}^{(\text{e})} = \mathcal{E}_\mathtt{p} \mathds{1}_\mathtt{q} \mathds{1}_\mathtt{s} \mathds{1}_\mathtt{t},
\end{align}
With the SVD specified, it is straightforward to confirm that $H_{\mathrm{plaq}}^{(\text{e})}$ can be diagonalized via
\begin{align}
    & H_{\mathrm{plaq}}^{(\text{e})} = \mathscr{U}^{(\text{e})\,\dagger}
    \big(Z_{\rm anc} \, \mathcal{S}^{(\text{e})} \big)\mathscr{U}^{(\text{e})},
\end{align}
with the diagonalizing transformation
\begin{align}
    &\mathscr{U}^{(\text{e})} = \mathsf{H}_{\rm anc} 
    \big(\ket{0}\!\!\bra{0}_{\rm anc} \mathscr{V}^{{(e)}\,\dagger} + \ket{1}\!\!\bra{1}_{\rm anc} \mathscr{W}^{{(e)}\,\dagger}\big)\mathscr{P}^{(\text{e})}.
\end{align}
A similar procedure brings $H_{\mathrm{plaq}}^{(\text{o})}$ to a diagonal form. Once again, all elements of the algorithm for $e^{-itH_{\mathrm{plaq}}^{(\text{e/o})}}=\mathscr{U}^{(\text{e/o})\,\dagger}
    e^{-itZ_{\rm anc} \, \mathcal{S}^{(\text{e/o})}}    \mathscr{U}^{(\text{e/o})}$ can be efficiently implemented using known quantum circuitry~\cite{Davoudi:2022xmb}.

The essential difference between our method compared with that used in Ref.~\cite{Kan:2021xfc} is that the latter entails splitting each and every link operator in the plaquette term and implementing each resulting term individually using known quantum circuits, whereas our method calls for splitting only a single link operator and implementing each resulting term individually. In other words, the method of Ref.~\cite{Kan:2021xfc} involves $2^4=16$ subevolutions per U(1)-plaquette operator, while ours calls for only two. Moreover, the excessive splitting as in Ref.~\cite{Kan:2021xfc} increases the number of noncommuting terms and hence the Trotter error.

\begin{figure}[t]
    \centering
    \includegraphics[width=0.99\linewidth]{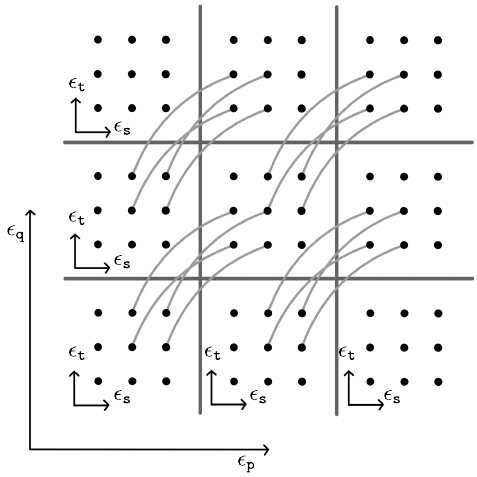}
    \caption{
    A schematic representation of the nonzero electric-basis matrix elements of the U(1) plaquette Hamiltonian, $H_{\mathrm{plaq}}=U_{\mathtt{p}} U_{\mathtt{q}} U_{\mathtt{s}}^\dagger U_{\mathtt{t}}^\dagger+\mathrm{H.c.}$, based on a 4D grid. Each point represents a basis state with quantum numbers $(\epsilon_{\mathtt{p}},\epsilon_{\mathtt{q}},\epsilon_{\mathtt{s}},\epsilon_{\mathtt{t}})$.
    $\epsilon_{\mathtt{p}}$ and $\epsilon_{\mathtt{q}}$
     ($\epsilon_{\mathtt{s}}$ and $\epsilon_{\mathtt{t}}$) are the coordinates along the coarse outer (fine inner) axes. Each edge represents a mixing between two states by $H_{\mathrm{plaq}}$, which generates moves of $\pm(1,1,-1,-1)$ in the grid. In this example, a truncation $\epsilon_i \in \{0,\pm1\}$ is in effect, hence three blocks of points along the outer axes and three points within each block along the inner axes.}
    \label{fig: 4D system and Hplaq graph}
\end{figure}

The above discussion can be illuminated by a graph-theoretic interpretation of the splitting of $H_{\mathrm{plaq}}$. First, one can associate basis states $\ket{\epsilon_{\mathtt{p}},\epsilon_{\mathtt{q}},\epsilon_{\mathtt{s}},\epsilon_{\mathtt{t}}}$ of a plaquette with points $(\epsilon_{\mathtt{p}},\epsilon_{\mathtt{q}},\epsilon_{\mathtt{s}},\epsilon_{\mathtt{t}})$ on a 4D grid. Second, one can connect two points $(\epsilon_{\mathtt{p}},\epsilon_{\mathtt{q}},\epsilon_{\mathtt{s}},\epsilon_{\mathtt{t}})$ and $(\epsilon_{\mathtt{p}}',\epsilon_{\mathtt{q}}',\epsilon_{\mathtt{s}}',\epsilon_{\mathtt{t}}')$ by an edge if the corresponding states are mixed by the action of $H_{\mathrm{plaq}}$, i.e.,
\begin{align}
    \bra{\epsilon_{\mathtt{p}}',\epsilon_{\mathtt{q}}',\epsilon_{\mathtt{s}}',\epsilon_{\mathtt{t}}'}H_{\mathrm{plaq}}\ket{\epsilon_{\mathtt{p}},\epsilon_{\mathtt{q}},\epsilon_{\mathtt{s}},\epsilon_{\mathtt{t}}}\neq 0.
\end{align}
Figure~\ref{fig: 4D system and Hplaq graph} defines a system for plotting the points of a 4D grid, which enables visualizing these mixings by $H_{\mathrm{plaq}}$. Two ``outer'' coordinates are used to specify $(\epsilon_{\mathtt{p}},\epsilon_{\mathtt{q}})$, while two ``inner'' coordinates are used to specify $(\epsilon_{\mathtt{s}},\epsilon_{\mathtt{t}})$. The points in Fig.~\ref{fig: 4D system and Hplaq graph} then represent the states of the plaquette if each link's electric field is truncated to three values, e.g., $\epsilon_i \in\{0,\pm 1\}$. The edges representing nonzero $H_{\mathrm{plaq}}$ matrix elements connect points that are related by moves of $\pm(1,1,-1,-1)$ in the space of electric quantum numbers.

Within Fig.~\ref{fig: 4D system and Hplaq graph}, one can observe chains that connect two states or three states; for higher cutoffs there are even longer chains of connected states. To simulate $H_{\mathrm{plaq}}$ directly, one would have to simultaneously diagonalize subspaces of all different dimensionalities up to the limit set by the cutoff. Splitting the Hamiltonian into multiple distinct terms can have the effect of breaking up the chains such that any individual subevolution under those terms only calls for rotating two-dimensional subspaces---rotations that can be analytically diagonalized in closed form.

\begin{figure*}[t]
    \centering
    \includegraphics[width=0.99\linewidth]{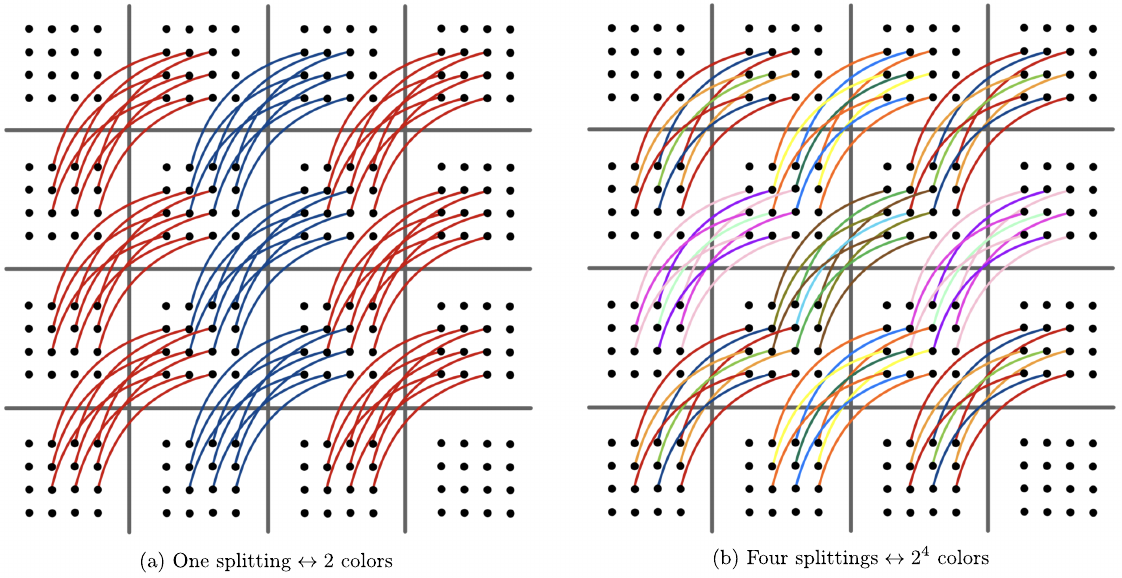}
    \caption{Colorings of a graph representing the truncated U(1) plaquette Hamiltonian, $H_{\mathrm{plaq}}$. In this example, each $\epsilon_i$ is truncated to four possible values to keep the drawing simple. The interpretation of the coordinates and edges is the same as in Fig.~\ref{fig: 4D system and Hplaq graph}.
    (a) A coloring of the graph using two colors, corresponding to the two subevolutions by $H_{\mathrm{plaq}}^{(\mathrm{e})}$ and $H_{\mathrm{plaq}}^{(\mathrm{o})}$.
    (b) A coloring of the graph using $2^4=16$ colors, corresponding to the 16 subevolutions called for by the method of Ref.~\cite{Kan:2021xfc}. In both (a) and (b), when restricting to any one color, there are no chains connecting more than two points, making it possible to analytically diagonalize each subevolution corresponding to each different color.
    }
    \label{fig:plaquette coloring}
\end{figure*}

In graph-theoretic terms, the edges of the graph associated with $H_{\mathrm{plaq}}$ can be colored with two colors. Figure \ref{fig:plaquette coloring}(a) depicts such a coloring when each $\epsilon_i$ is truncated to four possible values. One sees that, within the graph of any individual color, each point is connected to no more than one other point, corresponding to rotations of at most two-dimensional subspaces. Because of the two-color scheme, it should be possible to separate $H_{\mathrm{plaq}}$ into two ``sub-Hamiltonians'' that can each be separately diagonalized. This is exactly what $H_{\mathrm{plaq}}^{(\mathrm{e})}$ and $H_{\mathrm{plaq}}^{(\mathrm{o})}$ in Eq.~\eqref{eq:even-odd-H} correspond to, so they may be identified with the two colors of Fig.~\ref{fig:plaquette coloring}(a). In contrast, Fig.~\ref{fig:plaquette coloring}(b) depicts a coloring of the same graph using 16 colors, which is a valid but excessive coloring.
Each color can be mapped to one of the $2^4$ subevolutions that result from splitting each of the four quantum numbers, as called for by the method of Ref.~\cite{Kan:2021xfc}.

To connect back to the U(1) hopping term, recall that this term could be diagonalized without any splitting. The reason is, the analogous graph---a 3D graph in the coordinates of the fermion occupation numbers at sites $\mathtt{x}$ and $\mathtt{y}$ and the electric-field value on the link connecting them---can be colored with a single color, as discussed/depicted in Ref.~\cite{Stryker:2021asy}.

Table \ref{tab: summand counts} includes our count of $\Gamma=2$ for $H_{\mathrm{plaq}}$. While the factor of eight reduction using our method compared to Ref.~\cite{Kan:2021xfc} may seem modest, as it will become clear shortly, the redundant coloring, i.e., splitting, leads to an exponentially large number of subevolutions in $N_c^2$ for SU($N_c$) LGTs.
%

\section{SU(2) lattice gauge theory
\label{sec:SU(2)}
}
Thanks to the algorithm reviewed in the previous section for splitting Hamiltonians and diagonalizing the summands---which is a precursor to implementing any single subevolution within the PF decomposition of the Hamiltonian dynamics---the main concern becomes the number of proposed subevolutions entailed by the splitting. These summands exhibit a block-diagonal structure, and can therefore be fully diagonalized individually. Importantly, once these terms are identified, an SVD-based procedure similar to that presented for the U(1) case can be used to diagonalize them using basic quantum circuitry. The key differences that arise for SU($N_c$) are: i) The non-Abelian link operator is a ladder operator on multiple quantum numbers simultaneously. ii) The nonvanishing matrix elements of the non-Abelian link operator are nontrivial functions---rather than all equaling one, as was the case in Eq.~\eqref{eq:U(1)-link} for the U(1) link---and most of them are irrational, hence they demand more costly circuits to encode and evaluate. These properties are discussed and addressed in detail in Refs.~\cite{Davoudi:2022xmb,Kan:2021xfc}. In the following sections, we only focus on the summand-counting aspect of the algorithms.

We begin the discussion of non-Abelian LGTs with the SU(2) gauge group and consider coupling the gauge bosons to staggered fermions in the group's fundamental representation. An individual link is characterized by $N_c^2-1=3$ quantum numbers. The on-link Hilbert space is spanned by basis states of the form
\begin{align}
    &\ket{j,m^L,m^R} ,
\end{align}
where
\begin{align}
    &j \in \{0,\tfrac{1}{2}, 1, \cdots\}, \\
    &m^i \in \{-j,-j+1,\cdots j\},
\end{align}
with $i\in\{L,R\}$.
Above, $j$ labels an irrep of SU(2), and the azimuthal quantum numbers $m^L$ and $m^R$ are associated with rotations to the link-operator matrix from the left or right side. The above basis is the irrep or electric basis; it is to be contrasted with its conjugate group-element basis $\ket{ \theta_1 , \theta_2 , \theta_3 }$ wherein states are parametrized by coordinates in the SU(2) manifold (e.g., Euler angles of a rigid body).

In the fundamental (two-dimensional) representation, the link operator is a $2\times2$ matrix of operators $U_{\alpha,\beta}$ for $\alpha \in \{-\frac{1}{2},\frac{1}{2}\}$ and $ \beta \in \{-\frac{1}{2},\frac{1}{2}\}$. The link-operator matrix serves as a set of coordinate operators for the three-dimensional manifold of SU(2), and is conjugate to momentum-like operators that are the electric fields. The main property we are concerned with is the action of the link operator when expressed in the irrep basis:
\begin{align}
    &U_{\alpha,\beta} \ket{j , m^L , m^R}  = \nonumber \\  
    &\sum_{j' = \left| j - \frac{1}{2} \right| }^{j + \frac{1}{2}}
    C_{\alpha,\beta}(j,m^L,m^R;j')
    \ket{j' , m^L 
    +\alpha , m^R + \beta},
    \label{eq:SU(2)-link} 
\end{align}
with the coefficients $C_{\alpha,\beta}$ defined as
\begin{align}
    &C_{\alpha,\beta}(j,m^L,m^R;j') \coloneq  \nonumber\\
    &\sqrt{\tfrac{2j+1}{2j'+1}} \braket{
    j', m^L +\alpha | \tfrac{1}{2}, \alpha ; j, m^L 
    } \braket{ j', m^R + \beta | \tfrac{1}{2}, \beta ; j, m^R} .
    \label{eq:SU(2)-coefficients}
\end{align}
The exact form of coefficients $C_{\alpha,\beta}$ is irrelevant to the Hamiltonian-splitting procedure; the key information is which states appear in the right-hand side and the changes in quantum numbers.
The formulas for the $C_{\alpha,\beta}$ coefficients do, however, meaningfully impact the \emph{individual} cost of implementing any one subevolution.
 
According to Eq.~\eqref{eq:SU(2)-link}, applying $U_{\alpha, \beta}$ changes $m^L$ and $m^R$ by certain definite amounts---set by $\alpha$ and $\beta$---while $j$ can be incremented or decremented by $\tfrac{1}{2}$; i.e., there are terms for both $j'$ possibilities. To express these changes in the quantum numbers, it is convenient to define the following elementary ladder operators~\cite{Byrnes:2005qx}:
\begin{align}
\label{eq: Jpm action}
&J^{\pm} | j,m^L,m^R   \rangle = \left |  j\pm \tfrac{1}{2},m^L,m^R   \right \rangle, \\
&M_\alpha^L | j,m^L,m^R   \rangle =  |j, m^L+\alpha ,m^R  \rangle
\label{eq: M^L action}
\\
&M_\beta^R | j,m^L,m^R   \rangle =  |j, m^L ,m^R + \beta  \rangle.
\label{eq: M^R action}
\end{align}
In terms of the above operators, the link operator's components can be written as
\begin{align}
U_{\alpha,\beta} & =
M_{\alpha}^L M_{\beta}^R \left[ J^+ c_{1, \alpha}^L c_{1, \beta}^R  N_1 + J^- c_{2, \alpha}^L c_{2, \beta}^R  N_2 \right]
\nonumber\\
& \coloneq U_{\alpha,\beta}^{(+)} + U_{\alpha,\beta}^{(-)},
\label{eq:SU(2)-link-split}
\end{align}
where $N_k$ and $c_{k,\alpha}^{L/R}$ with $k \in \{1,2\}$ are shorthand for various known diagonal operators, whose form can be inferred from Eq.~\eqref{eq:SU(2)-coefficients} but their expressions will not be needed in the proceeding discussions. The form in Eq.~\eqref{eq:SU(2)-link-split} makes it clear that each distinct term in the link operator involves three ladder operators---one for each of the link's quantum numbers.

Recall that the gauge-matter interaction energy (hopping terms) and the magnetic energy (plaquette terms) are constructed from the link operators. These off-diagonal operators, with generally irrational coefficients, are harder to simulate when compared with the electric energy, which is diagonal in the irrep basis and has matrix elements that are quadratic forms. We next explain how to split the SU(2) hopping and plaquette terms into diagonalizable summands.

\subsection{Hopping terms}
The hopping Hamiltonian of a single link, analogous to Eq.~\eqref{eq:U(1)hopping-psi}, is
\begin{align}
    H_{\mathrm{hop}} &= \sum_{\alpha,\beta} \psi^\dagger_\alpha U_{\alpha,\beta}\psi_\beta ' + \mathrm{H.c.},
\end{align}
where, for brevity, we have omitted the link index $\mathtt{p}$ and the site index $\mathtt{x}$, and have denoted the operator at site $\mathtt{y}$ with a prime symbol.
For a fixed choice of $\alpha$ and $\beta$, one might observe that that $(\psi^\dagger_\alpha U_{\alpha,\beta}\psi_\beta ')^2=0$ and thus be inclined to choose that quantity as the $\mathscr{A}$ operator.
We do not proceed this way immediately because the SVD for $U_{\alpha,\beta}=U_{\alpha,\beta}^{(+)} + U_{\alpha,\beta}^{(-)}$ is nontrivial.
This is related to the fact that, in the graph-theoretic interpretation, the edges describing states connected by $\psi^\dagger_\alpha U_{\alpha,\beta}\psi_\beta '+\mathrm{H.c.}$ would require at least two colors---a consequence of the multiple possibilities for $j \to j\pm1/2$ induced by $U_{\alpha,\beta}$.
However, the graph of edges for $\psi^\dagger_\alpha U_{\alpha,\beta}^{(\sigma)}\psi_\beta '+\mathrm{H.c.}$ for either choice of $\sigma\in \{+,-\}$ can be colored with one color since either operator induces definite changes on all relevant quantum numbers.
It is also true that each choice of $\psi^\dagger_\alpha U_{\alpha,\beta}^{(\sigma)}\psi_\beta '$ squares to zero, resembles the ladder-operator content of the U(1) hopping term, and has an easily solvable SVD.
Thus, the counting of simulatable terms goes as follows: there are two choices for $\alpha$, two choices for $\beta$, and two choices for $\sigma$, for a total count of $2^3=8$ summands.
The count of eight is included in Table \ref{tab: summand counts}.

To diagonalize each of the distinct Hermitian operators $\psi^\dagger_\alpha U_{\alpha,\beta}^{(\pm)} \psi_\beta' + \text{H.c.}$, one proceeds according to the algorithms outlined for the U(1) case. The SVD matrices and transformations will be different compared to the U(1) case, but they can be still implemented with elementary circuit operations~\cite{Davoudi:2022xmb}. The diagonal operators after SVD are more costly to implement compared with the U(1) case due to the presence of nontrivial functions of number operators, i.e., $N_k$ and $c^{L/R}_{k}$ in Eq.~\eqref{eq:SU(2)-link-split}; nonetheless, the overall implementation cost remains polynomial in qubit number~\cite{Davoudi:2022xmb,Kan:2021xfc}. 

To summarize, our algorithm calls for the implementation of eight subevolutions per single-link hopping Hamiltonian, which is to be compared with the 64 subevolutions called for by the algorithm of Ref.~\cite{Kan:2021xfc}. The extra factor of eight arises from the unnecessary even-odd splitting of each of the three ladder operators $J^{\pm}$, $M_{\alpha}^{L}$ and $M_{\beta}^{R}$ in Eq.~\eqref{eq:SU(2)-link-split}. In the graph-theoretic representation, our algorithm amounts to using eight colors for the graph of edges connecting states in a seven-dimensional graph\footnote{We do not illustrate the graph of edges for the SU(2) hopping term or any later operators due to the high dimensionalities of the spaces in which they live.} (four dimensions for the two fermionic occupation numbers at two adjacent sites and three dimensions for the quantum numbers of the gauge link) while the algorithm of Ref.~\cite{Kan:2021xfc} corresponds to 64 colors.

\subsection{\label{sec: su(2) plaq counting}Plaquette terms}

A single plaquette's Hamiltonian has the form:
\begin{align}
    H_{\rm plaq}=
    2 \sum_{\alpha,\beta,\gamma, \delta} U_{\alpha,\beta} U_{\beta,\gamma} U_{\gamma,\delta} U_{\delta,\alpha},
\end{align}
(up to a constant multiplicative factor) where we have omitted the link labels and used the self-adjoint property of SU(2) traces to omit writing ``+H.c.''.
The task now is to identify summands to which one can apply the SVD algorithm.
One should expect a relatively large number of summands by considering the graph-coloring interpretation: for a given basis state of the four links, $H_{\mathrm{plaq}}$ can connect it to many different possible other basis states because of all the different choices for $\{\alpha,\beta,\gamma,\delta\}$ as well as possibilities for the transitions $j\to j\pm1/2$ on all four links.
Thus, a very large number of graph edges can connect to any one point in the plaquette's 12-dimensional grid of quantum numbers (with three coordinates associated with each of the four links).
To find the simulatable terms, we draw inspiration from the earlier examples by distilling the Hamiltonian down to summands that induce definite changes in all relevant quantum numbers. The counting of the terms proceeds as follows:
\begin{itemize}
    \item[$\diamond$] There are $2^4$ choices from the four indices $\{ \alpha,\beta,\gamma,\delta \}$ that can each take two values, and there are $2^4$ possibilities for the combinations of four $U^{(\pm)}$.
    \item[$\diamond$] The self-adjoint property of the SU(2) trace is exhibited by the observation that $$ U^{(\sigma_1)}_{\alpha,\beta} 
    U^{(\sigma_2)}_{\beta,\gamma} 
    U^{(\sigma_3)}_{\gamma,\delta} 
    U^{(\sigma_4)}_{\delta,\alpha} $$ (with $\sigma_i \in \{+,-\}$) induces the same change in quantum numbers as $$\bigl( U^{(-\sigma_1)}_{-{\alpha},-{\beta}} 
    U^{(-\sigma_2)}_{-{\beta},-{\gamma}} 
    U^{(-\sigma_3)}_{-{\gamma},-{\delta}} 
    U^{(-\sigma_4)}_{-{\delta},-{\alpha}} \bigr)^\dagger .$$ 
    In other words, the choice $(-{\alpha},-{\beta},-{\gamma},-{\delta},-\sigma_1,-\sigma_2 , - \sigma_3 , -\sigma_4 )$ is already accounted for by the choice of $(\alpha,\beta,\gamma,\delta,\sigma_1,\sigma_2,\sigma_3,\sigma_4)$ and Hermitian conjugation; a double counting arises if the choices are counted independently.
    Consequently, a factor of $\frac{1}{2}$ is needed to compensate for the overcounting. This gives a total of $2^4 \times 2^4 \times \tfrac{1}{2} = 2^7$ distinct rotational directions in the space of quantum numbers.
    \item[$\diamond$] Finally, in our method, within any single simulatable term, it is necessary to split at least one of the ladder-operator factors into odd and even parts that each square to zero.
    This gives an additional factor of 2, for a final count of $2^8=256$ subevolutions, as given in Table \ref{tab: summand counts}.
\end{itemize}

The counting outlined above is a notable improvement over that in Ref.~\cite{Kan:2021xfc} which amounts to 1,048,576 simulatable terms for a single plaquette Hamiltonian. Half of this count arises from double counting of the Hermitian-conjugate operators. The remaining count arises from splitting to even and odd parts all the three ladder operators in each distinct $U_{\alpha,\beta}^{(\pm)}$; there are four such links in each plaquette term, and there are $2^7$ total plaquette terms as noted above. These amount to $(2^3)^4 \times 2^7 = 524,288 $ terms to be diagonalized. Once again, such excessive splitting is unnecessary, and results in an increased call to diagonalization and diagonal-operator implementations, as well as increased Trotter error.

\section{SU(3) lattice gauge theory
\label{sec:SU(3)}
}
Next, we consider the SU(3) LGT in the Kogut-Susskind formulation with staggered fermions, whose continuum limit recovers QCD. For the SU(3) group, there are $N_c^2-1=8$ degrees of freedom per link. The on-link Hilbert space is spanned by irrep-basis states of the form
\begin{align}
    \ket{\omega , \lambda^L , \lambda^R } &\coloneq \ket{(p,q) , ( T^L, T^{L}_z, Y^L) , ( T^R, T^{R}_z, Y^R)},
\end{align}
where
\begin{align}
&p,q \in \left\{0,1,2,\cdots, \Lambda\right\},
\\
&T^{i} \in \left\{0, \frac{1}{2}, 1, \cdots, \frac{1}{2} (p+q)\right\},
\\
&T^{i}_z \in \left\{-\frac{p+q}{2} ,-\frac{p+q}{2} +\frac{1}{2},
\cdots,\frac{p+q}{2}\right\},
\\
&Y^{i} \in \left\{-\frac{q+2p}{3},-\frac{q+2p}{3}+\frac{1}{3},
\cdots,\frac{p+2q}{3}\right\},
\end{align}
with $i \in \{L,R\}$. Here, $\omega\coloneq(p,q)$ labels an irrep of SU(3) by two nonnegative integers and is analogous to the total angular momentum $j$ for SU(2).
The other quantum numbers are analogous to the azimuthal quantum numbers $m^{L}$ and $m^{R}$ in SU(2); they are labels for the SU(2)$\times$U(1) subgroup:
each side $i \in \{L,R\}$ of the link holds a total isospin $T^i$, third component of isospin $T^{i}_z$, and a U(1) hypercharge $Y^i$.\footnote{The quantum numbers $T$, $T_z$, and $Y$ are sometimes called ``magnetic'' quantum numbers for historical reasons. However, the irrep basis is an electric eigenbasis and the magnetic energy is strictly off-diagonal in this basis. For this reason, we will always refer to $T$, $T_z$, and $Y$ as the isospin and hypercharge quantum numbers, reserving ``magnetic'' for quantities directly associated with plaquette operators or their eigenbasis.}
This basis is to be contrasted with a conjugate, group-element basis parametrized by eight coordinates (i.e., generalized Euler angles) on the SU(3) manifold.

In the fundamental representation, the link operator is a $3\times3$ matrix of operators $U_{\nu ,\xi}$ for $\nu,\xi \in \{1,2,3\}$. $U$ transforms in the fundamental $(p,q)=(1,0)=\mathbf{3}$ irrep. Unlike SU(2), the matrix $U^\dagger$ is not unitarily equivalent to $U$, transforming instead in the antifundamental $(0,1)={\bf 3}^{* }$ irrep. We are mainly concerned with the action of the link operator in the irrep basis: 
\begin{widetext}
\begin{align}
    U_{\nu,\xi} & \ket{\omega, \lambda^L, \lambda^R} = 
    \sum_{\omega'}
    \sum_{{\lambda^L}'} \sum_{{\lambda^R}'} C_{\nu,\xi}(\omega, \lambda^L , \lambda^R ;
    \omega',{\lambda^L}' , {\lambda^R}' ) \ket{\omega', {\lambda^L}', {\lambda^R}'},
    \label{eq:SU(3)-link}
\end{align}
\end{widetext}
with coefficients $C_{\nu,\xi}$ given by
\begin{align}
    &C_{\nu,\xi}(\omega, \lambda^L , \lambda^R ;
    \omega', {\lambda^L}' , {\lambda^R}' )  =\sqrt{\frac{\dim (\omega)}{\dim (\omega')}} \times
    \nonumber\\
    & \hspace{1 cm} \braket{\omega', {\lambda^L}' | \omega, \lambda^L ; \mathbf{3} , \nu} \braket{\omega', {\lambda^R}' | \omega, \lambda^R ; \mathbf{3}, \xi} .
    \label{eq:SU(3)-coefficients}
\end{align}
The indices $\nu \in \{1,2,3\}$ and $\xi \in \{1,2,3\}$ are shorthand for the collections of, respectively, $(t^{L}, t^{L}_z ,y^{L})$ and $(t^{R}, t^{R}_z ,y^{R})$ quantum numbers associated with the components of $U$; labels `1', `2', and `3', correspond to $\left(\frac{1}{2}, \frac{1}{2}, \frac{1}{3}\right),\left(\frac{1}{2},-\frac{1}{2}, \frac{1}{3}\right)$, and $\left.(0,0,-\frac{2}{3}\right)$, respectively. The sum over $\omega^{\prime} \coloneq \left(p^{\prime},q^{\prime}\right)$ takes values in the set $\{(p+1, q),(p-1, q+1),(p, q-1)\}$, while the sums over ${\lambda^{i}}'=({T^i}',{T^{i}_z}',{Y^i}')$ take the values $(\left|T^{i}\pm t^{i}\right|,T^{i}_z + t^{i}_z,Y^i + y^i)$. $\operatorname{dim}(\omega)=\frac{1}{2}(p+1)(q+1)(p+q+2)$ and $\langle\omega^{\prime}, {\lambda^{L}}' | \omega, \lambda^{L} ; \mathbf{3}, \nu\rangle$ and $\langle\omega^{\prime}, {\lambda^{R}}' | \omega, \lambda^{R} ; \mathbf{3}, \xi\rangle$ are the SU(3) Clebsch-Gordan coefficients. To implement $U^\dagger$, one can take advantage of the identities that relate $U^\dagger$ to the $U$ operator in the antifundamental representation $\bm{3}^*$~\cite{deSwart:1963pdg,Byrnes:2005qx,Balaji:2025afl}.

To express distinct changes in the quantum numbers, one can define elementary ladder operators, analogous to the SU(2) case in Eqs.~\eqref{eq: Jpm action}-\eqref{eq: M^R action}:
\begin{alignat}{4}
    &P^{\pm} \ket{p} = \ket{p \pm 1} , & &Q^{\pm} \ket{q} = \ket{q \pm 1} , \\
    &{T^{i}}^\pm \ket{T^i} = \ket{T^i \pm \tfrac{1}{2}} , & \ \ \ &{T^{i}_z}^\pm \ket{T^{i}_z} = \ket{T^{i}_z \pm \tfrac{1}{2}} , \\
    &{Y^i}^\pm \ket{Y^i} = \ket{Y^i \pm \tfrac{1}{3}},
\end{alignat}
along with the composite ladder operators~\cite{Byrnes:2005qx}
\begin{align}
    &M_\nu^i = \left\{ \begin{array}{ll}
         {T^{i}_z}^+ {Y^{i}}^+ , & \nu = 1 \\
         {T^{i}_z}^- {Y^i}^{+} , & \nu = 2 \\
         ({Y^i}^{-})^2 , & \nu = 3
    \end{array} \right. \\
    &\mathcal{C}^{i}_{\omega,\nu} = \left\{ \begin{array}{ll}
         {T^i}^+ {D}^{i(a)}_{\omega,\nu} + {T^i}^- D^{i(b)}_{\omega,\nu} , & \nu \in \{ 1,2 \} \\
         D^{i}_{\omega\nu} , & \nu = 3
    \end{array} \right.
\end{align}
for $i \in \{L,R\}$. $D_{\omega,\nu}^{i}$ and $D_{\omega,\nu}^{i(a/b)}$ are diagonal operators in the irrep basis. The particular values of the diagonal elements are not relevant to the following discussion; they correspond to SU(3) Clebsch-Gordan coefficients~\cite{Kan:2021xfc} that are, as was noted, costly to evaluate. In terms of the above operators, the link operator's components can be written as
\begin{align}
    U_{\nu,\xi} &= M^L_\nu M^R_\xi \bigl[ P^+ \mathcal{C}_{1 \, \nu}^L \mathcal{C}_{1 \, \xi}^R N_1 + P^- Q^+ \mathcal{C}_{2 \, \nu}^L \mathcal{C}_{2 \, \xi}^R N_2 + \nonumber\\
    & \qquad \qquad \quad \ Q^- \mathcal{C}_{3 \, \nu}^L \mathcal{C}_{3 \, \xi}^R N_3 \bigr].
    \label{eq:SU(3)-link-ladder}
\end{align}
$N_k$ for $k \in \{1,2,3\}$ constitute yet another set of known diagonal operators in the irrep basis, whose forms do not matter in the proceeding discussions. The number and structure of transitions $U^\dagger$ induces is analogous to $U$, so we refer to only the $U$ form in all subsequent transition countings.

\subsection{Hopping terms}
The form of a single-link hopping Hamiltonian is $\psi^\dagger_\nu U_{\nu,\xi} \psi'_{\xi} + \mathrm{H.c.}$ In analogy to SU(2), any given element $U_{\nu,\xi}$ can be separated into terms that induce definite changes on the $(p,q)$ quantum numbers, of which there are three possibilities, as is evident from Eq.~\eqref{eq:SU(3)-link-ladder}. The number of transitions $U_{\nu,\xi}$ can induce on the isospin and hypercharge quantum numbers explicitly depends on the values of $\nu$ and $ \xi $. 
Specifically, the number of possible transitions depends on whether $\nu$ or $\xi$ is less than or equal to three:
\begin{itemize}
    \item[$\diamond$] The simplest case is $\nu = \xi = 3$, where $U_{3,3 }$ only induces definite transitions on the $Y^i$ quantum numbers in addition to the $(p,q)$ quantum numbers. Therefore, one has to simulate three terms, corresponding to the three possible changes in $(p,q)$ only.
    \item[$\diamond$] When exactly one of $\nu$ or $\xi$ is less than three, two possibilities for changing a $T^i$ quantum number arise (either raising or lowering it). This leads to $2 \times 3 = 6$ transitions to simulate for this particular choice of $U_{\nu,\xi }$. The total number of elements $U_{\nu,\xi }$ that fall into this category is four ($U_{1,3}$, $U_{2,3}$, $U_{3,1}$, and $U_{3,2}$), so in total, $6 \times 4=24$ terms are to be counted towards $\Gamma$.
    \item[$\diamond$] When $\nu < 3$ and $\xi < 3$, both $T^L$ and $T^R$ change, each with two possibilities. These two choices come with $2^2=4$ possibilities, while the possible choices of changes in $(p,q)$ are three as always. This leads to $4 \times 3 = 12$ transitions to simulate for this particular choice of $U_{\nu,\xi }$. The total number of elements $U_{\nu,\xi }$ that fall into this category is four ($U_{1,1}$, $U_{1,2}$, $U_{2,1}$, and $U_{2,2}$), so in total $12\times4=48$ should be counted towards $\Gamma$.
    \item[$\diamond$] As with the U(1) and SU(2) hopping terms, the presence of fermionic operators means no even-odd splitting of a bosonic ladder operator needs to be introduced. The final count is the sum of the previous steps (i.e., $3+24+48$), and amounts to a total of $\Gamma =75$ simulatable terms.
\end{itemize}

In contrast, Ref.~\cite{Kan:2021xfc} reports a count of 13,872 as an upper bound on the number of simulatable terms. This upper bound comes from splitting to even and odd parts up to four, six, and eight ladder operators in each of the 3, 24, and 48 terms counted above, respectively. This gives $2^4\times 3+2^6\times 24+2^8\times 48 =13,872$ total terms. Nonetheless, some terms are comprised of a product of fewer than four, six, or eight ladder operators. For example, consider the hopping term involving $U_{3,3}$. According to the right-hand side of Eq.~\eqref{eq:SU(3)-link-ladder}, $U_{3,3}$ is a sum of three terms, where the second term does have four ladder-operator factors but the first and third terms only have three.\footnote{The square of a ladder operator is counted as one ladder operator since it can be implemented as an incrementer of a single quantum number by two units.} When the exact number of ladder operators in each product is counted for each term, the total number of simulatable term via the splitting procedure of Ref.~\cite{Kan:2021xfc} is tightened to 9,248. This count is still much higher than the improved count of this work, $\Gamma=75$.

\subsection{\label{sec: su(3) plaq counting}Plaquette terms}
A single-plaquette Hamiltonian in the SU(3) LGT has the form
\begin{align}
\label{eq:su(3)-plaq}
H_{\rm plaq}&= U_{\alpha,\beta} U_{\beta,\gamma}  U^\dagger_{\gamma,\delta} U^\dagger_{\delta,\alpha}+ \mathrm{H.c.}
\end{align}
As with the link operator in the hopping term, one needs to account for the number of terms within any given element $U_{\nu,\xi}$ that induce distinct changes of quantum numbers, and this number depends on whether the indices are less than or equal to three. The counting of distinct terms contributing to Eq.~\eqref{eq:su(3)-plaq} goes as follows: 
\begin{itemize}
    \item[$\diamond$] The simplest term in the expanded trace is the one where all indices are equal to three, i.e., $U_{3,3} U_{3,3} U_{3,3} U_{3,3}$. Recall that $U_{3,3}$ for any given link changes the link's $Y^i$ quantum numbers by a definite amount, while there are three possibilities for how the $(p,q)$ quantum numbers may change. Hence, the total number of distinct changes that $U_{3,3} U_{3,3} U_{3,3} U_{3,3}$ can induce on the four-link plaquette is $3^4=81$.
    \item[$\diamond$] A maximally complicated term in the expanded trace is one where all of the indices are less than three, e.g., $U_{1,1} U_{1,1} U_{1,1} U_{1,1}$.
    For any given $U_{1,1}$, there are not only three possible changes in $(p,q)$, but also two possible changes in each of $T^L$ and $T^R$. The total number of choices for the given link is then $3\times2\times2=12$. The distinct choices for all four links will be $12^4=20,736$. There are $2^4$ combinations of $(\alpha,\beta,\gamma,\delta)$ that follow the counting of $U_{1,1}U_{1,1}U_{1,1}U_{1,1}$, amounting to $20,736 \times 2^4 = 331,776$ terms.
    \item[$\diamond$] The above examples inspire a derivation of the full count, including those given above, as follows: Collect the term multiplicities of the expanded $U_{\nu,\xi}$ into a matrix of their own:
    \begin{align}
        S &\coloneq 
        \left( \begin{array}{ccc}
             12 & 12 & 6 \\
             12 & 12 & 6 \\
             6 & 6 & 3
        \end{array} \right).
    \end{align}
    When the trace in Eq.~\eqref{eq:su(3)-plaq} is expanded out, the total number of distinct combinations of changes in quantum numbers will be $S_{\alpha,\beta} S_{\beta,\gamma} S_{\gamma,\delta} S_{\delta,\alpha} = 531,441$.
    \item[$\diamond$] As with the SU(2) plaquette, there is an additional factor of two needed for the splitting of a single bosonic ladder operator per 
    term, giving a final count of $\Gamma=1,062,882$.
\end{itemize}

In contrast, Ref.~\cite{Kan:2021xfc} reports the count $1,470,021,852,266,496$ as an upper bound on the number of simulatable terms. This count arises from splitting to even and odd parts up to eight, six, and four ladder operators in each of the three classes of link operators. In effect, one can replace the matrix $S$ above with
\begin{align}
    S' & \coloneq 
    \left( \begin{array}{ccc}
        12\times 2^8 & 12\times 2^8 & 6\times 2^6 \\
        12\times 2^8 & 12\times 2^8 & 6\times 2^6 \\
        6\times 2^6 & 6\times 2^6 & 3\times 2^4
    \end{array} \right).
\end{align}
The total number of terms is then $S'_{\alpha \beta} S'_{\beta \gamma} S'_{\gamma \delta} S'_{\delta \alpha} =1,470,021,852,266,496$. Since not all terms in each class have a maximum number of eight, six, or four terms, this count can be slightly improved, yielding a tightened count of $290,374,686,867,456$. This number is still significantly larger than our improved count reported above, that is $\Gamma$=1,062,882. This cost reduction can be expressed in terms of the number of colors, $N_c$: our strategy reduces the number of subevolutions to be implemented per Trotter step by a factor of $O(2^{4(N_c^2-1)})$ compared to Ref.~\cite{Kan:2021xfc}. This rough scaling can be understood by recalling that Ref.~\cite{Kan:2021xfc} assumes a worst-case scenario where each term in a given $U_{\nu,\xi}$ involves ladder operators for all $N_c^2-1$ quantum numbers on its link, hence one would have a splitting for each of $4\times(N_c^2-1)$ quantum numbers around the plaquette; by contrast, our algorithm splits a term involving any number of ladder-operator factors into just two simulatable terms.

We conclude this section by translating the above improvement in the number of simulatable terms into a reduction of the simulation cost of QCD within a (second-order) PF algorithm. First, one notes that the total time-evolution gate cost is the cost of a single Trotter step times the number of Trotter steps, $s$, see Eq.~\eqref{eq:V2}. The cost of a single Trotter step is dominated by the cost of plaquette operators: As already observed, the number of simulatable terms is much larger for the plaquette Hamiltonian than for the hopping Hamiltonian in the irrep basis. Moreover, the plaquette Hamiltonian has a large norm toward the continuum limit, i.e., as the gauge-matter coupling decreases.\footnote{We suppressed the constant factors in Hamiltonian terms in this papers, but they can be restored by consulting Refs.~\cite{Byrnes:2005qx,Kan:2021xfc}.} One plaquette operator involves $\Gamma$ subroutines (oracle calls), where $\Gamma$ is the number of simulatable terms. With either method (Ref.~\cite{Kan:2021xfc} or ours), a single subroutine involves similar arithmetic operations and can be carried out to similar precision, such that the costs of these subroutines are similar [$O(1)$ different], regardless of which subevolution is being considered and which decomposition strategy has been chosen. The number of Trotter steps, on the other hand, scales with $\alpha_2^{1/2}$ according to Eq.~\eqref{eq:p2-formula}, with $\alpha_2$ being the norm of a function of the second-order nested commutators of the Hamiltonian terms. As argued in Ref.~\cite{Kan:2021xfc}, the scaling of $\alpha_2$ can be bounded by $O(\Gamma)$, so that $s$ scales as $\Gamma^{1/2}$.\footnote{In Ref.~\cite{Davoudi:2022xmb}, the upper bound used for $\alpha_2$ was set at $O(\Gamma^3)$ instead of $O(\Gamma)$. Nonetheless, the argument of Ref.~\cite{Kan:2021xfc} may apply to the upper bounding there as well to reduce the simulation cost of the SU(2) LGT.} Putting all these together gives an overall scaling of $\Gamma \times \Gamma^{1/2} = \Gamma^{3/2}$ for the second-order PF simulation cost. Thus, for the SU(3) plaquette, with our $\Gamma$ being smaller by a factor of $\mathcal{R} \coloneq 1,470,021,852,266,496/1,062,882 \approx 1.38 \times 10^9$ relative to Ref.~\cite{Kan:2021xfc}, the gate cost falls by a factor of
\begin{align}
\mathcal{R}^{3/2}=(1.38 \times 10^9)^{3/2} \approx 10^{13.7}.
\end{align}
%

\section{Other algorithms
\label{sec:others}
}
There are other approaches to simulating QCD. In order to enable an apples-to-apples comparison of the associated cost, an end-to-end analysis of the algorithms in the same setting has to be performed and the precision targets have to be comparable.

Besides the complete analysis by Kan and Nam using PFs, another error-bounded analysis was presented by Rhodes, Kreshchuk, and Pathak~\cite{Rhodes:2024zbr} using near-optimal, post-Trotter time-evolution algorithms, which we outline in this section. This approach achieves an exponentially improved scaling in $N_c^2$ compared to the PF algorithm by Kan and Nam~\cite{Kan:2021xfc}.

We also discuss in this section the approach of Ciavarella, Klco, and Savage~\cite{Ciavarella:2021nmj}, which was later refined in Ref.~\cite{Balaji:2025afl}. While lacking the full error-bounded simulation-cost analysis, this approach relies on eliminating gauge redundancy in the electric (irrep) basis using physical constraints. A question that arises is whether such an approach asymptotically performs better than the algorithms based on redundant encodings, such as that presented in this work and in Refs.~\cite{Kan:2021xfc,Rhodes:2024zbr}---a question we partly address in this section.

Finally, we note that our comparative study in this section is limited to quantum simulation of QCD in the truncated electric basis. Over the past decade, progress is reported using other approaches as well, such as large-$N_c$ approximations~\cite{Ciavarella:2024fzw,Ciavarella:2025bsg,Modi:2026syn}, maximally gauged-fixed formulations~\cite{Froland:2025bqf,Grabowska:2024emw}, orbifolds~\cite{Bergner:2024qjl,Halimeh:2024bth,Halimeh:2025ivn}, group-element representations~\cite{Lamm:2019bik,Haase:2020kaj,Bauer:2021gek} and discrete subgroups~\cite{Lamm:2024jnl,Gustafson:2022xdt,Perez:2025cxl}, q-deformed and similar representations~\cite{Zache:2023dko,Rhodes:2026atz,Webb-Mack:2026bkg}, and simulations in the Coulomb gauge~\cite{Li:2024ide,Yao:2025uxz}. Once a full simulation-cost analysis is achieved for these approaches using the system setting and parameters reported here, one can begin to compare their costs relative to the methods discussed in this paper.

\subsection{Simulation with near-optimal algorithms}
The approach of this work results in nearly 14 orders of magnitude reduction in the simulation cost of the QCD Hamiltonian within the Kogut and Susskind formulation compared with Ref.~\cite{Kan:2021xfc}. However, the number of subevolutions to be simulated in each Trotter step, $\Gamma$, remains large. The number of terms is dominated by the magnetic interactions in the electric-field basis, yielding $\Gamma \sim 10^6$ with the decomposition of this work. Recall that the total cost goes as $\Gamma^{3/2}$ using a second-order PF, contributing a factor of $\sim 10^9$ to the T-gate complexity. This cost motivates considering approaches that bypass the need for decomposition of the exponentiated Hamiltonian. One such approach is based on near-optimal simulation algorithms rooted in block encodings of the Hamiltonian. In fact, Rhodes, Kreshchuk, and Pathak present~\cite{Rhodes:2024zbr} a resource analysis of the simulation cost of QCD within the Kogut-Susskind framework, which reports a $\sim 10^{23} - 10^{27}$-fold improvement in the T-gate cost compared to the PF approach of Ref.~\cite{Kan:2021xfc} (for a range of system sizes and parameters toward continuum and infinite-volume limits). Therefore, the result of this work, while an improvement in the cost of PF-based simulations, remains at least nine orders of magnitude higher than that reported by Ref.~\cite{Rhodes:2024zbr}. In the following, we briefly describe the building blocks of the algorithm of Ref.~\cite{Rhodes:2024zbr} and attempt to track the sources of such a large cost reduction.

Consider a $d$-dimensional lattice with the total number of sites $N^d$. The goal is to simulate a geometrically-local LGT Hamiltonian $H$ for time $T$ with error $\epsilon$, starting from a state with bounded local quantum numbers at any gauge link. For SU(3), the initial-state bound is such that $p,q \leq \Lambda_0$ at all links. The algorithm of Ref.~\cite{Rhodes:2024zbr} uses the Haah, Hastings, Kothari, and Low (HHKL) method~\cite{Haah:2018ekc}, which decomposes the full time-evolution operator into time evolution of local blocks $\mathcal{B}$. Simulation of each block only requires evolution for time $t = O(1)$. The simulation of the system then requires $O(T)$ such time segments and $O(N^d)$ such blocks.

To simulate each block and each segment, Ref.~\cite{Rhodes:2024zbr} then uses a qubitization algorithm~\cite{Low:2016znh}. Qubitization---an algorithm that is based on block encoding of the Hamiltonian and functions of it---has a time complexity proportional to the norm of the Hamiltonian. Since the electric-field term in the Kogut-Susskind Hamiltonian is unbounded (or realistically bounded with a large cutoff), it is advantageous to simulate physics in the interaction picture with respect to the electric-field term~\cite{Rajput:2021khs,Tong:2021rfv}. 

Specifically, $\ket{\psi(t)}=e^{-itH}\ket{\psi(0)}=e^{-itH_E}\mathcal{T}[e^{-i\int^t dt' H_I(t')}]\ket{\psi(0)}_I$, where $H_I(t)=e^{-itH_E}(H_\text{hop}+H_M+H_\text{plaq})e^{itH_E}$. Here, $\mathcal{T}$ denotes the time-ordering operator, and $H_E$, $H_\text{hop}$, $H_M$, and $H_\text{plaq}$ are the electric, hopping, mass, and plaquette Hamiltonians, respectively. The dynamics of the electric-field term $e^{-itH^\mathcal{B}_E}$ within each block of the HHKL algorithm can be fast forwarded, as it is diagonal in the electric-field basis. The time-dependent Hamiltonian $H_I(t)$ can be simulated using a truncated Dyson-series approach~\cite{Berry:2014ivo,Low:2018pte}. The total asymptotic gate cost of these steps, as shown in Ref.~\cite{Rhodes:2024zbr}, turns out to be ${O}\big(N^dT\text{polylog}(N^dT \Lambda_0 \epsilon^{-1})\big)$. This is to be compared with the second-order PF algorithm asymptotic cost, which for the SU(3) LGT goes as $O\big(N^{3d/2}T^{3/2}\Lambda \epsilon^{-1/2}\text{polylog}\big(N^{3d/2}T^{3/2}\Lambda \epsilon^{-3/2} \big)\big)$~\cite{Kan:2021xfc}.

A key question is whether the algorithm of Ref.~\cite{Rhodes:2024zbr} harbors a large prefactor associated with the number of simulatable blocks. This question is ultimately related to the scaling of the algorithm as a function of $N_c$. To address this question, one first notes that Ref.~\cite{Rhodes:2024zbr} employs a sparse-access oracle as a precursor to qubitization. The oracle first finds and enumerates the nonzero states achieved by the action of the Hamiltonian terms (via a subroutine called $O_F$); it then evaluates the associated amplitudes which depend on the Clebsch-Gordan coefficients (via a subroutine called $O_H$). 

For the costly plaquette operator, the subroutine $O_F$ consists of two successive components. First, a unary-iteration circuit selects a plaquette and one of the $N_c^4$ possible color contractions arising from the trace. The registers encoding the four color indices are then used as controls for a second circuit that encodes the action of the corresponding color-resolved plaquette (corresponding to the 4-tuple of color indices) on the gauge-field registers. Since the selection circuit and the plaquette-encoding circuit are applied sequentially within a single call to $O_F$, their gate costs are added together (rather than multiplied). The plaquette encoding itself factorizes into four link updates, each of which depends only on the pair of color indices associated with that link.
The controlled implementation of each link-operator factor must, therefore, be defined for each of the $N_c^2$ possible input pairs of color indices. For a fixed component $(\nu,\xi)$ of a given link, the action of $U_{\nu,\xi}$ produces up to 12 nonzero candidate states for SU(3), whose construction requires a total of 88 controlled signed-incrementer/decrementer operations. The combined cost of implementing all four link-operator factors is, therefore, proportional to $4N_c^2 \times C(N_c)$ where $C(N_c)$ is an $N_c$-dependent function [88 for SU(3)]. The subroutine $O_H$, on the other hand, evaluates the amplitudes, with a cost that at most is proportional to $4 \times 12$, with the same origin for each factor as above (with no additional $N_c^2$ factor because the color pair is supplied as control/input information to the amplitude-evaluation circuit $O_H$).

The block-encoding cost is determined by two calls each to $O_F$ and $O_H$, together with controlled rotations. This block encoding is then queried by the qubitization algorithm in the interaction picture, with a number of queries that scale with the norm of the interaction-picture Hamiltonian and the simulation time. This yields the claimed $N_c^4$ scaling from the $N_c^4$ possible color contractions of the plaquette operator in $O_F$. Two comments are, however, in order. First, the $N_c^2$-dependent link-update contribution to $O_F$ is larger than the $N_c^4$ color-selection contribution for SU(3); moreover, the general $N_c$ dependence of the fixed-color-link-update and amplitude-evaluation circuits has not been established. Second, a sparse-access block encoding generally carries a normalization that depends on the matrix sparsity~\cite{gilyen2019quantum}. Consequently, the multiplicative sparsity of the non-Abelian plaquette operator may enter the block-encoding normalization---and hence the number of queries to the block encoding---even though it does not appear explicitly in the gate costs quoted for $O_F$ and $O_H$.

This difference in $N_c$ scaling is significant compared to the result of Ref.~\cite{Kan:2021xfc}, which has at least a $2^{4(N_c^2-1)}$ scaling. Our term-count prefactor, which removes this exponential factor from the cost of Ref.~\cite{Kan:2021xfc}, is still a factor of $10^3$ higher compared to Ref.~\cite{Rhodes:2024zbr}. Moreover, the full second-order Trotterization cost is proportional to $\Gamma^{3/2}$, which brings the cost difference associated with the number of terms between our algorithm and that in Ref.~\cite{Rhodes:2024zbr} to a factor of $10^6$. 

One may wonder what additional factors contribute to the remaining cost difference (i.e., an extra factor of around $10^3-10^7$ for various parameter sets considered in Ref.~\cite{Rhodes:2024zbr}). The exponential improvement in $\epsilon^{-1}$ is not significant except for ultra-precise simulations. Similarly, the optimal time scaling [$O(T)$ vs. $O(T^{3/2})$ for the second-order PF] is only important for long simulations. The use of the HHKL algorithm brings down the cost of naive qubitization from $O(V^2)$ to $O(V\text{polylog}(V))$, which is asymptotically more optimal than the second-order PF volume scaling $O(V^{3/2})$. However, achieving the HHKL scaling requires the size of the HHKL blocks to be mach smaller than the lattice volume~\cite{Haah:2018ekc}, which is yet to be confirmed for SU($N_c$) LGTs. Other differences include the use of a local encoding of the fermions in Ref.~\cite{Rhodes:2024zbr} compared with the nonlocal encoding of Ref.~\cite{Kan:2021xfc}, and the HHKL approach for simulating time dynamics in Ref.~\cite{Rhodes:2024zbr}. Both of these features can be applied to the approach of Ref.~\cite{Kan:2021xfc} (and hence of this work) to potentially improve the cost. 

It is likely that the Trotter-error bounds can be improved, e.g., by obtaining more realistic but hard to compute empirical estimates of the errors (using approaches such as randomized compiling~\cite{Childs:2019atb,Campbell:2019fez}) or by applying constraints imposed by the initial states~\cite{Heyl:2019rtj,Sahinoglu:2020dwp,Yi:2021cxm,Gong:2023hbt,Burgarth:2023cal,Su:2020gzf,Watson:2023oov,Mizuta:2025aoi}, including states' available quantum resources~\cite{Zhao:2024xtl,Zhang:2026cbs}. Such improvements may further reduce the cost difference between PF-based and block-encoding-based algorithms of QCD in the electric-field basis. More research is needed to scrutinize both types of algorithms to fully understand the origin of cost difference between the two methods.

\subsection{Simulation in the local-multiplet basis with classically precomputed rotations}
\noindent 
In this work, as well as in Refs.~\cite{Kan:2021xfc,Rhodes:2024zbr}, the Kogut-Susskind SU($N_c)$ LGT in the irrep basis is encoded by allocating to each of the $N_c^2-1$ bosonic quantum numbers a qubit register of size $\eta$ (where $\eta \propto \log(\Lambda)$ for the binary encoding). Nonetheless, the gauge-invariant Hilbert is much smaller than the full Hilbert space, suggesting that a smaller set of quantum numbers may be at play.

This guidance, and the analogy to an earlier SU(2) analysis~\cite{Klco:2019evd}, led Ciavarella, Klco, and Savage in Ref.~\cite{Ciavarella:2021nmj} to rely on classical pre-processing to explicitly construct the gauge-invariant states at vertices of the lattice and to identify all the physical transitions among them. The subevolutions associated with these transitions are subsequently encoded in a quantum circuit. A gauge-invariant state is formed by summing over states with any possible gauge-field quantum numbers, weighted by SU(3) Clebsch-Gordan coefficients to ensure that the irreps overall add up to a gauge singlet. Since the Gauss's law operators at a vertex all commute with the quadratic and cubic Casimirs of any attached link, one may characterize a gauge-singlet state with the irreps $(p,q)$ of the involved links. Not all assignments of $(p,q)$ to the links are consistent with the formation of a gauge singlet; the physical states are those that have the allowed $(p,q)$ combinations.

The evaluation of SU(3) Clebsch-Gordan coefficients is an old problem that is more complicated than its SU(2) counterpart. The reason is, when two irreps of SU(2), i.e., angular momenta $j_1$ and $j_2$, are added together, a total angular momentum $j_3$ can be formed in a unique way. In an SU(2) LGT, this means that gauge-singlet configurations of trivalent vertices may be completely characterized by the angular momenta (or Casimir quantum numbers) of the attached links. In the representation theory of SU(3), on the other hand, the tensor product $(p_1,q_1) \otimes (p_2,q_2)$ may contain multiple copies of an irrep $(p_3,q_3)$.\footnote{The most commonly cited example is the tensor product of two octets, which can combine to form a third octet in two distinct ways (referred to as ``symmetric'' and ``antisymmetric''). In general, there is no agreed-upon way to handle this new degree of freedom.} This means that $(p,q)$ labels are insufficient to completely characterize all states, and additional multiplicity indices\footnote{Denoted by $\Gamma$ in Refs.~\cite{Ciavarella:2021nmj,Balaji:2025afl}.} are needed at the vertices to account for all gauge-invariant states. 

Once the states have been constructed, one must evaluate the Hamiltonian matrix elements. The most complex matrix elements correspond to a single plaquette operator. The cost of applying the plaquette operator on any given basis state is not expected to scale dramatically with increasing the irrep quantum numbers $(p,q)$ because the link operators can only change the irreps in a few ways, as discussed in Sec.~\ref{sec:SU(3)}. Once the state resulting from the action of the link operators are obtained, one has to take inner products of this state with the set of gauge-invariant states. This process yields a Hamiltonian matrix that is indexed by the gauge-invariant quantum numbers: irreps $(p,q)$ for the links as well as the sites' multiplicity indices.

The plaquette operators' action, in principle, does not depend on the states of the external links, as it is entirely determined by the actions on individual links comprising the plaquette, see Eq.~\eqref{eq:SU(3)-link}. The irrep quantum numbers of the four on-plaquette links are in general consistent with a variety of different irrep configurations of the external links. Nonetheless, when a plaquette operator is applied to a gauge-invariant state, the changes in irreps along the four on-plaquette links are accompanied by changes that affect the Clebsch-Gordan coefficients, as needed to maintain gauge singlets at each corner of the plaquette. These Clebsch-Gordan coefficients, that are functions of the external fluxes meeting at the corners, therefore, create a dependence on the external fluxes. This has led to the notion of (externally) \emph{controlled plaquette operators}, i.e., operators that update the fluxes of on-plaquette links with amplitudes that are determined by the states of externally attached links. A given update to the fluxes of on-plaquette links is in general compatible with many different possible configurations of the external links, leading to a number of transition amplitudes that increases for higher cutoffs. In other words, by eliminating all quantum numbers but $(p,q)$ (and possible multiplicity indices), the plaquette operator's action becomes less local.

In the context of quantum simulation, the above repackaging of plaquette operators has been implemented in a Trotterized time-evolution scheme. In this scheme, a plaquette operator is simulated by a PF decomposition of its own, involving individual subevolutions for each and every independent matrix element of the plaquette operator. These subevolutions are Givens rotations on the plaquette's links that are logically controlled by the irreps of the external links. The logical controls are nontrivial quantum gates. Thus, a key cost metric for simulation in this scheme is the number of nonzero amplitudes contained by a plaquette operator, which determines the number of PF subevolutions.

To understand the scaling of the number of nonzero matrix elements, let us first ignore the Gauss's law constraints as well as the multiplicity indices.
First, the number of possible states of the plaquette, as characterized by the $(p,q)$ pairs on the four links, goes as $\Lambda^8$. The number of possible states one basis state can transition to from another does not depend on $\Lambda$ because of the limited transitions possible, as mentioned above, rendering the $\Lambda^8 \times \Lambda^8$ transition matrix to be sparse. Thus, the number of nonzero entries would scale with the number of columns only.

Now there is an additional complication, which is that the matrix elements actually depend on the $(p,q)$ pairs of external links as well, as mentioned earlier. Hence, there must be a distinct matrix for each and every possible external configuration of gauge-invariant quantum numbers. The scaling of the number of these ``external-control sectors'' is $(\Lambda^{(N_c-1)(v-2)})^4$ for plaquettes with vertices with valency $v$. In other words, the full plaquette acting on the local-multiplet basis is a block-diagonal matrix involving $(\Lambda^{(N_c-1)(v-2)})^4$ independent blocks, each with at most $\Lambda^8$ nonzero matrix elements (for a square or cubic lattice).

\begin{figure}[t!]
    \centering
    \includegraphics[width=0.47\linewidth]{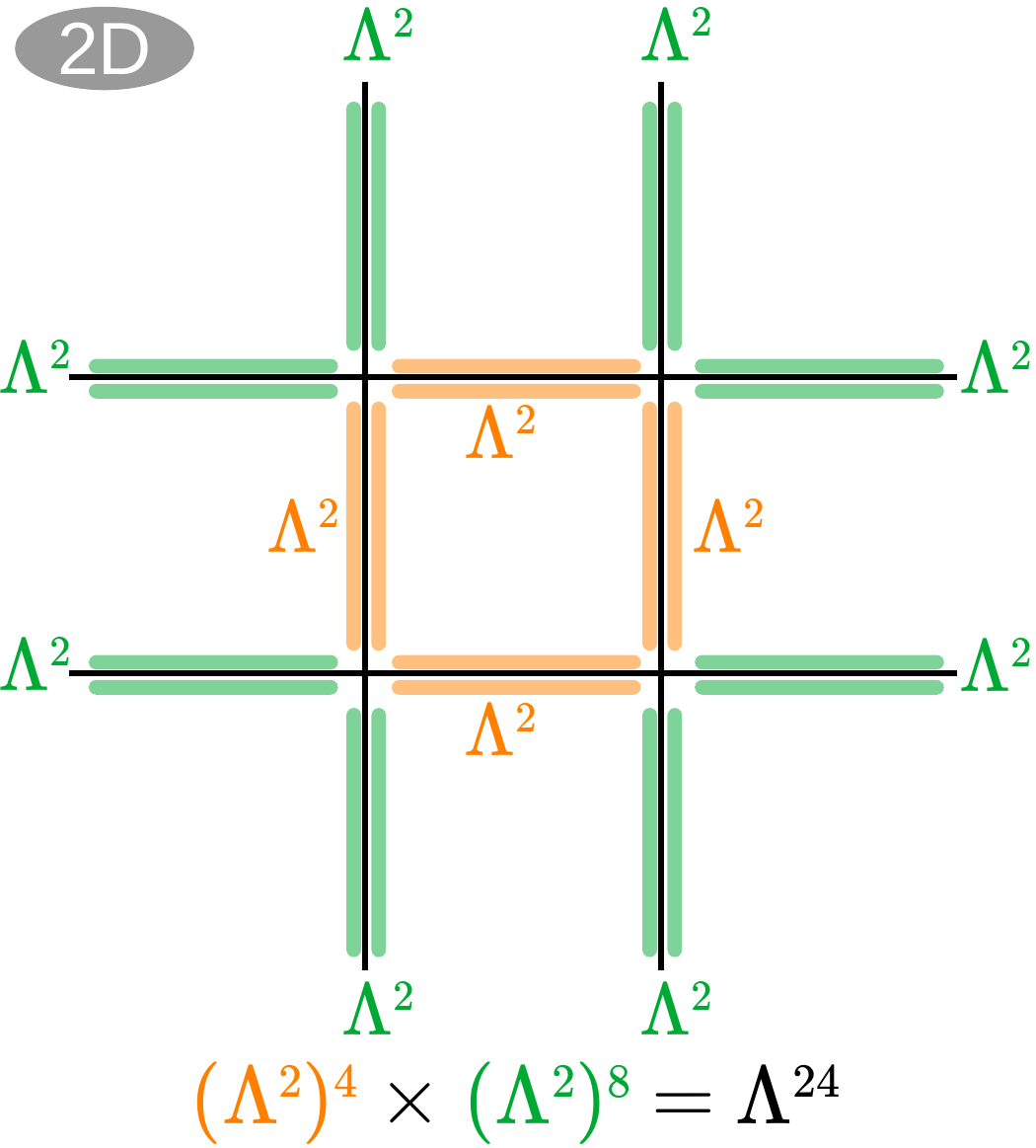}\hspace{0.04\linewidth}
    \includegraphics[width=0.47\linewidth]{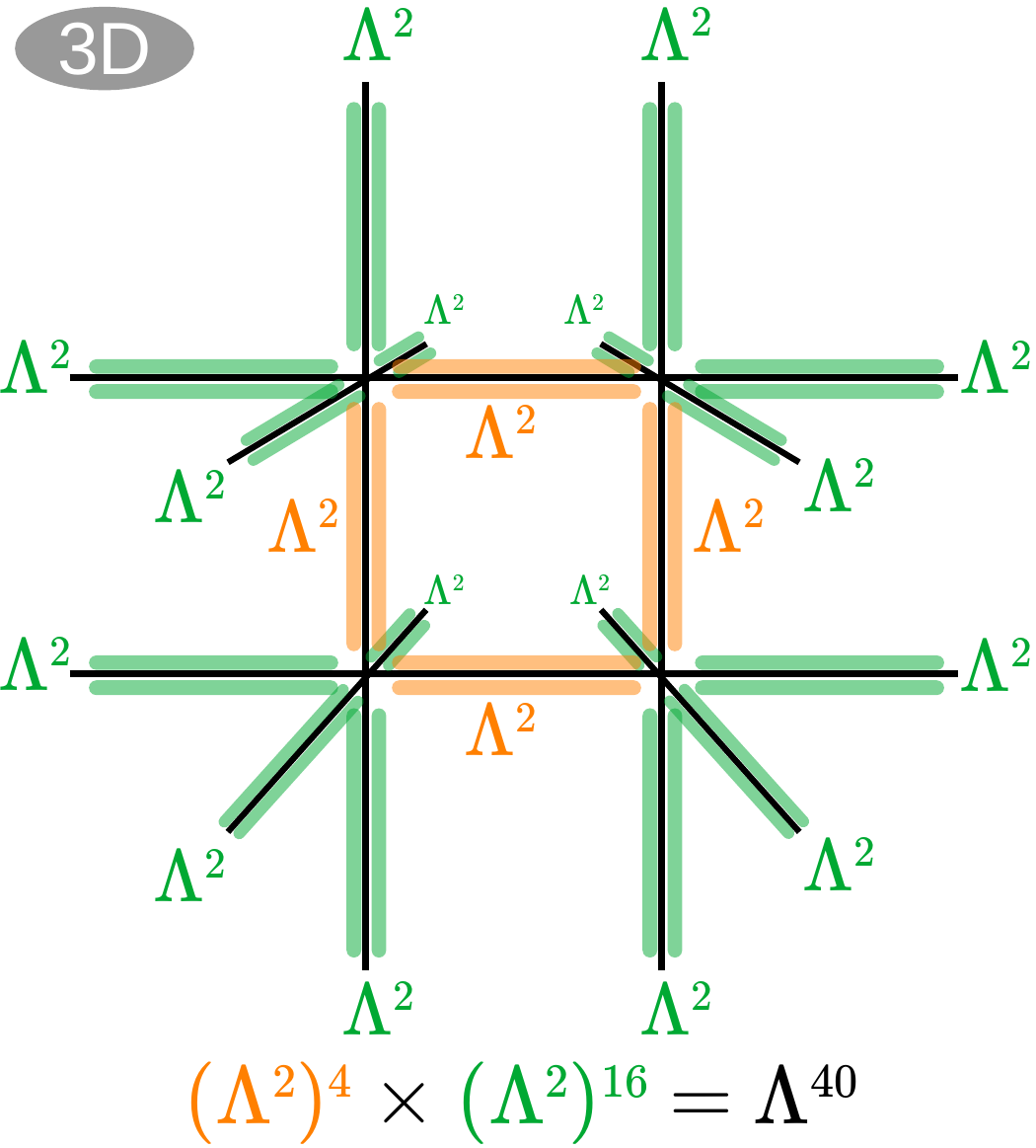}
    \caption{
    The estimated scaling of the number of nonzero matrix elements of the plaquette operator, for square (left) and cubic (right) lattices, when working in the local-multiplet basis~\cite{Ciavarella:2022zhe}. Orange links on the plaquette contribute $\Lambda^8$ scaling from the number of $(p,q)$ configurations the four links can be in. Green links are external to the plaquette but contribute additional powers of $\Lambda$ due to the functional dependence of the matrix elements on the $(p,q)$ configurations of these external links.
    }
    \label{fig:local-irreps}
\end{figure}

Finally, one should consider whether or not the Gauss's law constraints or multiplicity indices change the scaling argument. Gauss's laws constraints must have the effect of reduce the absolute number of nonzero transition-matrix entries, but multiplicity indices would have the effect of increasing them. Apparently, as argued in Ref.~\cite{Ciavarella:2022zhe}, the scaling behavior empirically appears to still approximately follow the  polynomial scaling outlined above, so that the above counting is still a good proxy for the scaling. Putting the above observations all together, in the ideal case of trivalent vertices, the scaling of the plaquette is $\Lambda^8$; unfortunately, however, one cannot tile 2D or 3D space using exclusively trivalent vertices without introducing plaquettes that have more than four links. For the standard case of a square (cubic) lattice, in which all vertices are $v=4$ ($v=6$), the scaling is $\Lambda^{24}$ ($\Lambda^{40}$). This scaling argument is summarized in Fig.~\ref{fig:local-irreps}.

To achieve a full quantum-simulation cost, one needs to take a step further and evaluate the Trotter error associated with the splitting of the plaquette operator to many separate Givens rotations. 
While operations belonging to the different external-control sectors commute, those within each sector may not commute. This source of error has not yet been accounted for, hindering a full cost comparison with the results considered in this work. A key difference, nonetheless, is the large polynomial scaling of the algorithm with $\Lambda$ compared with the modest ${O}\big(\Lambda\,\text{polylog}(\Lambda)\big)$ scaling of our algorithm and the algorithm of Refs.~\cite{Kan:2021xfc,Rhodes:2024zbr}. It appears that the cost saving in the number of qubits [from $O(8\eta)$ down to at least $O(2\eta)$] is overshadowed by a significant growth in the circuit complexity. Moreover, the classical pre-processing that sets up the quantum circuits can be expensive for large cutoff values required toward the continuum limit.

Further refinement of the algorithm of Ref.~\cite{Ciavarella:2021nmj} is possible. For example, Balaji et al.~\cite{Balaji:2025afl} offer a range of optimizations in implementing each plaquette transition operator, adopting strategies such as pruning, gray ordering, and fusions. Their algorithm, nonetheless, still scales with the number of physical transitions of the plaquette, which is unaltered by such improvements. The fusion strategy, which combines appropriate entangling operations into a smaller set, can reduce the number of gates, but all physical transitions still need to be pre-processed and encoded in a circuit before further optimizations are sought. Moreover, the gate reductions appear to not be asymptotically large, leaving the large scaling with the cutoff as a bottleneck of the algorithm in the far term.

\section{Conclusions
\label{sec:outlook}
}
\noindent
In this work, we demonstrate a substantial reduction in the gate count of Trotterized time evolution under the quantum-chromodynamics Hamiltonian within the Kogut-Susskind lattice formulation in the truncated electric basis. The reduction is achieved by an application of the singular-value decomposition algorithm of Ref.~\cite{Davoudi:2022xmb}, which yields a term decomposition with a number of terms that is exponentially smaller in $N_c^2$;
for U(1), SU(2), and SU(3) gauge theories, the number of costly magnetic-term subevolutions are reduced by factors of $8$, $4.096 \times 10^3$, and $1.383052730 \times 10^9$, respectively, when compared with the previous state-of-the-art~\cite{Kan:2021xfc}. Using a second-order product formula, these term reductions translate to nearly $20$-fold, $10^5$-fold, and $10^{14}$-fold reductions, respectively, in the simulations' T-gate counts. The gate-count reductions hold regardless of simulation parameters such as system size, simulation time, and electric-field truncation cutoff. 

Moreover, our work provides a comparison of simulation strategies, full cost, and asymptotic cost scaling between this work and two other leading results using Kogut-Susskind lattice Hamiltonian in the truncated electric basis. First, the near-optimal, post-Trotter approach of Ref.~\cite{Rhodes:2024zbr} yields a factor of $10^3$ reduction in the number of terms and a factor of $10^6$ in the overall associated cost compared with our work, regardless of simulation parameters, along with additional reductions associated with optimal scaling in time and inverse error, and likely due to the use of a local evolution structure and fermion-to-qubit encoding. Second, the local-multiplet-basis approach of Ref.~\cite{Ciavarella:2021nmj} results in a compact encoding of the local degrees of freedom in terms of the Casimir quantum numbers, but it also introduces controlled-plaquette building blocks that leads to an algorithmic complexity that scales as $\Lambda^{24}$ for 2D square and $\Lambda^{40}$ for 3D cubic lattices. This is to be contrasted with algorithms of Refs.~\cite{Kan:2021xfc,Davoudi:2022xmb}, and of this work, which scale at most as ${O}\big(\Lambda\,\text{polylog}(\Lambda)\big)$.

The result of this work demonstrates that improving the cost of quantum simulating QCD is a continuous effort; as new understanding in formulations, algorithms, and cost analysis emerge, the QCD algorithmic cost can be re-analyzed with new tools, and significant improvements are possible. The initial high cost estimates and unfavorable asymptotic scalings should not discourage the progress. One should also bear in mind that present-day QCD computations using classical computing resources are themselves resource-intensive, requiring fractions of exascale supercomputers per year per calculation, which amounts to $\sim 10^{25}$ floating-point operations. It is conceivable that precision QCD simulations would demand the use of quantum supercomputers in the future.

On the path to further improvement in QCD quantum-resource analysis, one must account for a number of issues and the correlated interplay among them:
\begin{itemize}
  \item[$\diamond$] Boson-truncation errors in the electric basis within the Kogut-Susskind theory are often controlled by exponentially suppressed bounds of Tong et al.~\cite{Tong:2021rfv}~and subsequent improvements~\cite{Peng:2023bzl}; nonetheless, these bounds become polynomial in the cutoff in the continuum limit~\cite{Tong:2021rfv}. Truncation errors in other formulations and representations need to be rigorously bounded or empirically estimated. Progress along these lines continues~\cite{Yang:2026zpa,Ciavarella:2025tdl}.
  \item[$\diamond$] The continuum limit of lattice gauge theories, via a simultaneous extrapolation to the zero lattice spacing and infinite truncation cutoff is important and needs to be done in parallel to time-digitization or other time-evolution schemes to balance the systematics and avoid unnecessarily large resource estimates. Such analysis in the context of PF and post-Trotter algorithms for lattice gauge theories are studied in recent years~\cite{Carena:2021ltu,Kane:2025ybw,Carena:2022kpg,Ciavarella:2023mfc,Illa:2025njz}, but need to be applied systematically to other formulations and representations.
  \item[$\diamond$] Finite-volume and finite-time-extent systematics will also be important to be accounted for in simulation-cost analysis; progress in quantifying such infrared effects has been reported in recent years for Minkowski-time quantum-field-theory simulations~\cite{Briceno:2020rar,Burbano:2025pef}.
  \item[$\diamond$] Last but not least, the analysis of this work and other works reported here often considers solely the time-evolution task in a quantum computer. The reason is that rigorous algorithmic analysis are possible for various time-evolution algorithms (although they may be loosely bounded, like the Trotter approaches). Preparation of nontrivial states in QCD is an important task and would require more dedicated resource studies within different approaches. On the same footing, the observable estimation and measurements are subject to errors that need to be accounted for in an end-to-end analysis of the QCD simulation cost.
\end{itemize}

The above discussions make it clear that quantum-algorithm developments for gauge theories in general, and for QCD in particular, remain a frontier research problem and a complete picture of required quantum resources will likely emerge in the coming years.

\vspace{1 cm}

\noindent
\emph{Note:} Close to the submission of this manuscript, a manuscript by Draper was released~\cite{Draper:2026bcj}, in which the near-optimal algorithm of Ref.~\cite{Rhodes:2024zbr} is combined with the precomputed local-multiplet-basis approach of Ref.~\cite{Ciavarella:2021nmj}, upon various improvements in both methods. Future work can compare this algorithm's performance with those presented here.

\section*{Acknowledgments
\label{sec:acks}
}
\noindent
We acknowledge early contributions by Alexander Shaw, and fruitful discussions with Anthony Ciavarella, Christopher Kane, Natalie Klco, Michael Kreshchuk, and Mason Rhodes. Z.D. was supported by the U.S. Department of Energy (DOE), Office of Science, Office of Nuclear Physics (award no. DE-SC0026067); the DOE, Office of Science, Office of Advanced Scientific Computing Research (ASCR), program in Accelerated Research in Quantum Computing, Fundamental Algorithmic Research toward Quantum Utility (FARQu); the U.S. National Science Foundation’s Quantum Leap Challenge Institute (award no. OMA-2120757); and Maryland Center for Fundamental Physics, Department of Physics, and College of Computer, Mathematical, and Natural Sciences at the University of Maryland. Z.D. is further grateful to Ignacio Cirac and the Max Planck Institute for Quantum Optics (MPQ), Garching, Germany, for their hospitality during the completion of this work. The visit to MPQ was made possible by a Humboldt Research Fellowship by the Alexander von Humboldt Foundation. J.R.S. was supported by the U.S. DOE, Office of Science (award no. DE-AC02-05CH11231), partially through the Quantum-Information-Science-Enabled Discovery (QuantISED) program for High Energy Physics (KA2401032). 

\bibliography{bibi.bib}
\end{document}